\documentclass[twocolumn]{aastex631}

\newcommand{\tess}{\textit{TESS} }
\newcommand\ntargets{59} 
\newcommand\ntois{76} 
\newcommand{\nPCs}{25} 
\newcommand{\nCPs}{51} 
\newcommand{\brg}{Br$\gamma$}

\newcommand{\nStartotal}{86} 
\newcommand{\nplanetsottal}{142} 

\begin{document}

\title{Uniform Metallicity Measurements of M Dwarf Planet Hosts Support Metallicity-Dependent Sub-Neptune Formation}

\author[0000-0002-1845-2617]{Emma V. Turtelboom}\altaffiliation{William $\&$ 
Caroline Herschel Postdoctoral Fellow}\affiliation{Department of Physics \& Astronomy, McMaster University, 1280 Main Street West, Hamilton, Ontario, Canada}\affiliation{Department of Astronomy, 501 Campbell Hall, University of California, Berkeley, CA 94720, USA}
\author[0000-0002-8965-3969]{Steven Giacalone}\altaffiliation{NSF Astronomy and Astrophysics Postdoctoral Fellow} \affiliation{Department of Astronomy, California Institute of Technology, Pasadena, CA 91125, USA}
\author[0009-0000-7274-7523]{Rebecca Gore} \affiliation{Department of Physics \& Astronomy, San Francisco State University, 1600 Holloway Avenue, San Francisco, CA 94132, USA}
\author[0000-0001-8189-0233]{Courtney D. Dressing}\affiliation{Department of Astronomy, 501 Campbell Hall, University of California, Berkeley, CA 94720, USA}
\author[0000-0001-5383-9393]{Ryan Cloutier}\affiliation{Department of Physics \& Astronomy, McMaster University, 1280 Main Street West, Hamilton, Ontario, Canada}
\author[0000-0002-6825-351X]{Karina Kimani-Stewart}\affiliation{Department of Physics \& Astronomy, Georgia State University, 25 Park Place, Atlanta, GA 30303, USA} \affiliation{RECONS Institute, Chambersburg, PA 17201, USA}
\author[0000-0003-0012-9093]{Aida Behmard} \affiliation{Center for Computational Astrophysics, Flatiron Institute, 162 Fifth Ave, New York, NY 10010, USA}
\author[0000-0002-9548-1526]{C.M. Lisse} \affiliation{Johns Hopkins University Applied Physics Laboratory, 11100 Johns Hopkins Road, Laurel, MD 20723, USA}
\author[0000-0003-1799-1755] {M.L. Sitko} \affiliation{Center for Exoplanetary Systems, Space Science Institute, 4750 Walnut Street, Suite 205, Boulder, CO 80301, USA}

\correspondingauthor{Emma V. Turtelboom}
\email{eturtelboom@berkeley.edu}

\begin{abstract}
M dwarfs are the most common sites of planet formation in the Milky Way. Planet occurrence and composition are closely linked with the availability of metals in protoplanetary disks, which can be probed by measuring the metallicity of planet host stars. In this work, we measure the metallicities ([M/H] and [Fe/H]) of \ntargets{} M dwarfs hosting \ntois{} planets and candidates using medium-resolution near-infrared spectra collected with IRTF/SpeX. We combine these results with literature metallicity measurements for planet-hosting cool dwarfs, and present a sample of \nStartotal{} stars hosting \nplanetsottal{} candidate, validated, and confirmed planets with homogeneously derived stellar parameters. Using our updated stellar radii, we calculate planet radii from TESS transit depths for both the confirmed (N = \nCPs{}, 0.6 - 12.5 $\rm{R_\oplus}$, median $\rm{R_p} = 1.8 \rm{R_\oplus}$) and candidate (N = \nPCs{}, 0.6 - 7.2 $\rm{R_\oplus}$, median $\rm{R_p} = 2.1 \rm{R_\oplus}$) planets. We compare the metallicity distributions of super-Earth and sub-Neptune host stars, finding that M dwarfs hosting sub-Neptunes are statistically more metal-rich than those hosting super-Earths. This result is robust to the radius valley prescription used, and is likely not due to differences in the stellar samples considered. This result supports the hypothesized formation pathway whereby sub-Neptunes form beyond the water ice line where they can accrete volatiles before migrating inwards to their observed locations. The enhanced inventories of refractory elements throughout the disk and of volatiles beyond the ice line in metal-rich disks around low-mass stars may contribute to the preference seen in the observed planet sample for sub-Neptunes to orbit metal-rich M dwarfs.
\end{abstract}


\section{Introduction} \label{sec:intro}
Our understanding of the more than 6000 exoplanets discovered to date is highly dependent on our understanding of the stars that they orbit. Despite the growing sample of directly imaged planets around young stars \citep[e.g.][]{lagrange+09, macintosh+15}, the majority of exoplanets orbit older stars and are discovered using transit and radial velocity (RV) observations. These indirect detection methods infer the presence and parameters of a planet through changes in host star observables. As a result, imprecise or inaccurate stellar mass, radius, effective temperature, surface gravity, and metallicity values lead to inaccurate planet parameters and incorrect inferences about planetary system formation. Previous work \citep[e.g.][]{petigura20, macdougall+23} has demonstrated that reducing uncertainties in stellar parameters reveal planet demographic features in sharper detail, enabling more precise connections between observed planets and planet formation mechanisms. 

Metallicity and mass are two fundamental characteristics of stars that impact planet formation, as they reflect the total mass, lifetime, and metallicity of the disks in which planets form \citep{Williams2011,Ribas2015}. The metal content of these disks sets the solid material available to form protoplanet cores, and thus stellar metallicities are commonly used as a proxy for disk dust-to-gas ratios. Small planets ($<4 R_\oplus$) are more common around low-mass stars than around Sun-like stars \citep{Dressing2013,  Hardegree-Ullman2019, Ment2023, gillis2026}. On the other hand, the occurrence of short-period (P $< 10$ d) giant planets, known as Hot Jupiters, appears to peak around solar-mass stars \citep{beleznay+kunimoto22, bryant+23}. The occurrence of giant planets is highly correlated with stellar metallicity \citep{santos01, fischer+valenti05, ghezzi+18}. A metal-rich disk increases the chance that a protoplanet reaches the critical mass to trigger runaway gas accretion before the gas disk dissipates, resulting in a strong occurrence-metallicity relationship for gas giants. 

On the other hand, the relationship between small planet occurrence and stellar metallicity is much more uncertain. \citet{buchave+12} and \citet{Kutra2021} found that the formation of small planets may be agnostic to stellar metallicity, while \citet{Lu2020} report an increase in planet occurrence for planets with radii between 2 and 5 $\rm{R_\oplus}$ as stellar metallicity increases. Small extrasolar planets, with radii between that of the Earth and of Neptune, are not found in the Solar System, but are very common in the Milky Way \citep{Howard2012}. Furthermore, the occurrence of small planets increases around low-mass stars, which are themselves the most common stars in the Milky Way \citep{Henry2024}. Therefore, understanding how small planets around low-mass stars form and evolve is critical to forming a full picture of planet formation in the solar neighbourhood and beyond.

With the large sample of observed planets, it is possible to compare and characterize planet sub-populations. For example, small planets can be further divided into super-Earths ($\rm{R_p} \lesssim 1.8 R_\oplus$) and sub-Neptunes ($2 R_\oplus \lesssim \rm{R_p} \lesssim 4 R_\oplus$). These two groups are separated by the radius valley, first posited to exist following theoretical arguments \citep{Owen2013} and later observed in radial velocity survey results \citep{Fulton2017}. The dominant physical processes governing the radius valley remain unclear. Super-Earths and sub-Neptunes may form as distinct populations \citep[e.g.][]{Lee2022, Lopez2018, Chiang2013,Dawson2015, MacDonald2020}, may be only distinguished by post-formation atmospheric mass-loss \citep[e.g.][]{Bean2021}, or may be formed through a combination of these two scenarios \citep{Burn2024}. 

Constraining planet interiors to reveal the nature of the radius valley is challenging due to degeneracies inherent in inferring complex interior structures from few observables: planet mass, radius, and instellation. In one scenario, sub-Neptunes are enriched in volatiles \citep[e.g.][]{Leger2004, Rogers2015, Mousis2020, Burn2024} while super-Earths are dry and rocky \citep[e.g.][]{Rogers2025}. Alternatively, sub-Neptunes may include carbon-rich planets \citep{linseager25}, ice-rock mixtures \citep{Kovacevic2022}, and volatile-poor planets \citep{Chen2025}. If super-Earths and sub-Neptunes form in different disk locations, these environmental differences may also be evident in their host stars \citep{Brugger2017, Owen2018}. Characterizing planet host stars therefore offers another window into the physical properties underlying the observed radius valley.

Previous investigations of the radius valley around low-mass stars found that as stellar mass decreases super-Earths become an order of magnitude more common than sub-Neptunes \cite{Ment2023, gillis2026}. Furthermore, the slope of radius valley separating the two populations decreases in magnitude in period-radius space \citep{gupta2022} and the radius valley moves to higher planet instellation \citep{Parashivamurthy2025, gillis2026} with decreasing stellar mass. There remain, however, clear sub-populations in density space \citep{Luque2022}, suggesting that compositionally distinct super-Earths and sub-Neptunes around low-mass stars had unique formation pathways.\citet{Ho2024} and \citet{Venturini2024} suggest that the filling in of the radius valley around low-mass stars may point to a dispersion in core-mass fractions due to icy planets or steam worlds, although this scenario is currently indistinguishable from increased photo-evaporative or collisional mass loss for planets around late-type stars. 

The metal contents of stars, and specifically of low-mass stars, can be measured in several ways \citep{Passegger2022}. By combining model stellar atmospheres and radiative transfer codes, synthesized spectra can be compared to observations \citep[e.g.][]{Bean2006, Wanderley2025}. However, this process of spectral synthesis is complicated for M dwarfs due to the high volume of complex molecular features and lack of continuum at optical wavelengths. Several other methods thus make use of supervised machine learning \citep[e.g.][]{Antoniadis-Karnavas2024, Behmard2025} and photometric calibrations using FGK-M binary systems \citep[e.g.][]{Bonfils2005,Duque-Arribas2023}. Near-infrared (NIR) spectra are commonly used to characterize low-mass stars \citep[e.g.][]{Hejazi2015, Mann13b, MannAN}, due to strong metallicity-sensitive features in this wavelength regime as well as enhanced luminosity of cool dwarfs at longer wavelengths. Nevertheless, there remain systematic offsets between different metallicity measurement methods \citep[e.g.][]{Jofre2019, Pass2025}, highlighting the need for uniformly derived metallicities to make inferences about the planet populations orbiting low-mass stars.

In this paper, we used IRTF/SpeX NIR spectra to characterize \ntargets{} cool dwarf stars observed by \tess{} which host confirmed planets or planet candidates. In Section \ref{sec:sample}, we present our sample and target selection criteria, and we present the observations used in this work in Section \ref{subsec:obs}. We then detail the methods used to measure metallicities and other key stellar parameters from our observations in Section \ref{sec:methods}. We provide updated planet parameters and vet the \nPCs{} planet candidates in our sample in Section \ref{sec:planets}. Finally, in Section \ref{sec:discussion} we discuss the relationship between stellar metallicity and planet parameters, and conclude in Section \ref{sec:conclusions}.

\section{Sample} \label{sec:sample}

We selected the targets in our sample from the sample of \tess Objects of Interest (TOIs) released and vetted by the \tess Science Office\footnote{Available at \url{https://exofop.ipac.caltech.edu/tess/}}, with a view to target cool dwarfs with transiting planet candidates or confirmed planets. We first restricted our sample to targets with effective temperatures $<4000 K$ and stellar radii $<0.75 R_\odot$, in order to target cool stars in the M and late-K spectral classes. These initial stellar parameters were reported in the \tess{} Input Catalog \citep[TIC v8.2][]{ticv82}. For each star, the stellar radius was calculated using \textit{Gaia} parallaxes and magnitudes. Stellar effective temperatures were calculated using an empirical photometric relation with colour \citep{Stassun2018}. We restricted the sample to targets observable using the IRTF ($-30^\circ < \delta < +65^\circ$). We then removed fainter stars ($\rm{K mag} > 13$). We also excluded targets for which the planet candidate radius reported on ExoFOP was greater than $1 R_J$; deeper transits are more likely to be False Positives caused by eclipsing binaries. Finally, we prioritized obtaining observations for stars hosting multiple planets or planet candidates and for brighter stars. Our final sample of \ntargets{} stars is presented in Table \ref{tab:updated_params}. 

Within this sample, there are \nCPs{} confirmed planets and \nPCs{} planet candidates. These planets are quite diverse; they range from the size of Venus to that of Jupiter, and their orbital periods range from 0.5 to 48.7 days. Additionally, eight of the planets have measured masses, six of which have radii $< 4 \rm{R_\oplus}$. These planets span a range of densities, from 1.8 $\pm 0.3 \rm{g/cm^3}$ \citep[GJ 436 b,][]{Maciejewski2014b} to 7.0 $\pm 1.2 \rm{g/cm^3}$ \citep[GJ 486 b,][]{Trifonov2021a}. Of the \ntois{} planets and planet candidates, 67 have radii $< 4 \rm{R_\oplus}$, and 2 have radii similar or greater than that of Saturn\footnote{\url{https://nssdc.gsfc.nasa.gov/planetary/factsheet/saturnfact.html}} ($9.1 \rm{R_\oplus}$).

\begin{figure*}
    \centering
    \includegraphics[width=\textwidth]{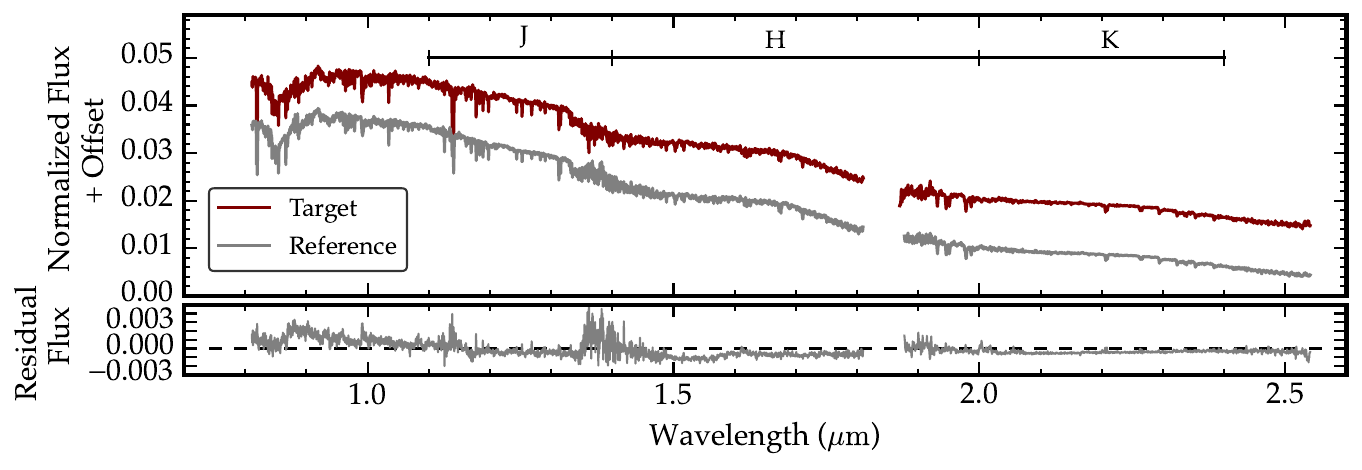}
    \caption{The observed spectrum for TIC 219041246 (TOI-5713, red) compared with a IRTF standard spectrum of Gl 273 (grey). The spectra are offset for visual clarity. The wavelength ranges of the J, H, and K spectral band-passes are indicated for reference. The bottom panel shows the residuals between the observed and standard spectra. The regions of larger residual scatter near 1.4 $\mu m$ are due to high $\rm{H_2O}$ telluric contamination in those wavelength regimes.}
    \label{fig:spectra}
\end{figure*}

\section{Observations} \label{subsec:obs}
\subsection{IRTF/SpeX} \label{subsec:irtfobs}

We collected medium-resolution spectra for our targets using the SpeX instrument \citep{rayner+03} on the 3m NASA Infrared Telescope Facility (IRTF) between May 2023 and January 2024. We collected observations using the SXD mode with a $0.3'' \times 15''$ slit, providing a spectral resolution of R $\approx$ 2000 over a wavelength range of 0.7 $-$2.55 $\mu m$ \citep{rayner+04}. We used exposure times of no more than 180 seconds, and we collected 65 spectra of \ntargets{} unique targets.

For each observation, we used an ABBA nod pattern, with the A and B positions on the slit separated by $7\farcs5$. In this pattern, each pair of images is subtracted, in order to remove the impact of dark and other telescope currents on the images. This approach does not correct for sky background, as this varies between locations A and B. However, we synced the slit to the parallactic angle in order to ensure that atmospheric dispersion occurred along the slit, minimizing light loss. We repeated the ABBA nod pattern with the goal of achieving a S/N of at least 100 per resolution element. 

We also acquired observations of A0V stars to use for telluric corrections. We identified the nearest appropriate A0V star to each science target using the SpeX locator tool\footnote{Available at \url{https://irtfweb.ifa.hawaii.edu/~spex/find_a0v/}}. Where possible, we observed A0V stars directly following science observations, and no less than every two hours. We observed standard stars with angular distances and airmass differences no greater than $15^\circ$ and 0.1, respectively.

After each observation of a science target, or set of science observations in a given area of the sky, we took calibration images using the internal quartz and Thorium-Argon lamp. These images were executed using the IRTF/SpeX calibration macro associated with the SXD observation mode and $0\farcs3$ slit length. 

We reduced our spectra using the \texttt{Spextool} reduction pipeline \citep{spextool}. \texttt{Spextool} was developed specifically for IRTF/SpeX data, and performs three main steps: calibration frame preparation, spectral extraction, and post-extraction processing of images. Telluric contamination in our spectra was removed using our observations of A0V stars in between science targets and using \texttt{xtellcor} \citep{spextool2}. We used the calibration and A star images taken nearest to each science target for reduction purposes. We include a representative spectrum in Figure \ref{fig:spectra}.

\subsection{Imaging} \label{subsec:imagingobs}
We incorporate existing imaging observations of 38 of our targets when vetting planets in Section \ref{sec:validation}, in order to rule out nearby eclipsing binaries as false positive scenarios. We describe these observations in the Appendix \ref{sec:appendix_imaging}. 

\subsection{Photometry} \label{sec:dataphot}
In order to characterize the planetary systems around our targets, specifically for the \nPCs{} targets which host planet candidates (rather than confirmed and/or validated planets), we used photometric observations collected by the Transiting Exoplanet Satellite Survey \citep[TESS,][]{Ricker2015}. We used the two-minute cadence light curves produced by the TESS Science Processing Operations Center \citep[SPOC,][]{Jenkins2016}. The number of sectors observed per target ranged from 1 to 42, with a median of 6 sectors of TESS data per target with a median baseline of 5.2 years. 

\section{Stellar Characterization}\label{sec:methods}

\subsection{Spectroscopic}
We followed the methodology of \citet{dressing+19} and \citet{Gore2024} to measure the stellar masses, radii, effective temperatures, and metallicities of our targets. We first used the Python implementation\footnote{\url{https://github.com/stevengiacalone/tellrv}} of the \texttt{tellrv} package \citep{newton+14, tellrv} to calculate the absolute radial velocity for each target. We analyzed the J, H, and K bands of each spectrum separately, and cross-correlated each spectrum with the corresponding telluric-corrected standard star (A0V) spectrum. We then calculated the radial velocity of each target by taking the median of the J, H, and K band radial velocities, and shift the science spectrum by this median radial velocity.

Using these shifted lab-frame spectra, we calculated the iron abundance relative to hydrogen ([Fe/H]) and the total metal abundance relative to hydrogen ([M/H]). We calculated these metallicities using the relationships for cool dwarfs (spectral types K7 - M5 for -1.04 $<$
[Fe/H] $<$ +0.56) presented in \citet{Mann2013a}. These empirical relations were derived using a sample of 112 wide binary systems with a solar-type primary and cool dwarf secondary. We specifically calculated these metallicities in the $J$ and $K$-bands of our spectra using the metal-sensitive features centered at 2.2079, 2.3242 and 2.3844$\mu {\rm m}$ identified by \citet{Mann2013a}. \citet{Onehag2012} found that J-band derived metallicities are reliable indicators of the metal content of M dwarfs. However, \citet{dressing+19} argue that K-band metallicities are most reliable for IRTF/SpeX spectra due to increased telluric contamination in the other bands, and we therefore report the $K$-band-derived metallicities and adopt these values for the remainder of our analysis. The resulting metallicity measurements have a typical uncertainty of 0.1 dex, and are reported in Table \ref{tab:updated_params} We measure a range of iron abundances ([Fe/H]: [-0.57, +0.57] dex, median [Fe/H] = -0.03 dex) and of total metallicity ([M/H]: [-0.48, +0.42] dex, median [M/H] = -0.04 dex). 

\subsection{Photometric} \label{sec:methods_phot}
We also calculated the stellar luminosity following the methodology in \citet{dressing+19}. Firstly, we found the r-band magnitude in the Carlsberg Meridian Catalogue \citep{muinos+evans14}, and the V and J magnitudes in the TIC \citep{ticv82}. Secondly, we calculated the V-J bolometric correction formula from \citet{mann+15}, incorporating our measured metallicity (specifically, the spectroscopic [Fe/H] value) to improve the accuracy of the relation. We added an additional error of 0.012 magnitudes in quadrature to the bolometric correction error to account for scatter in the empirical calculation, as recommended by \citet{mann+15}. Thirdly, we used distances from the TIC to calculate the absolute J-band magnitude ($M_J$), from which we calculated the stellar luminosity using the bolometric correction.

In order to measure the stellar radius, we used the empirical $M_K$-radius relation presented in \citet{mann+15}, again incorporating our measured [Fe/H] values. We calculated the absolute K magnitude ($M_K$) for each target using the stellar distance and K magnitude values reported in the TIC. We then calculated the stellar radius, adding an error of $2.7\%$ in quadrature to reflect the scatter in the empirical relation presented in \citet{mann+15}. 

We calculated stellar mass using the mass-magnitude relation presented in \citet{Mann19}, adding a systematic uncertainty of 0.02 $M_\odot$ in quadrature to reflect the scatter in this empirical relation. 

\begin{figure*}
    \includegraphics[width=\textwidth]{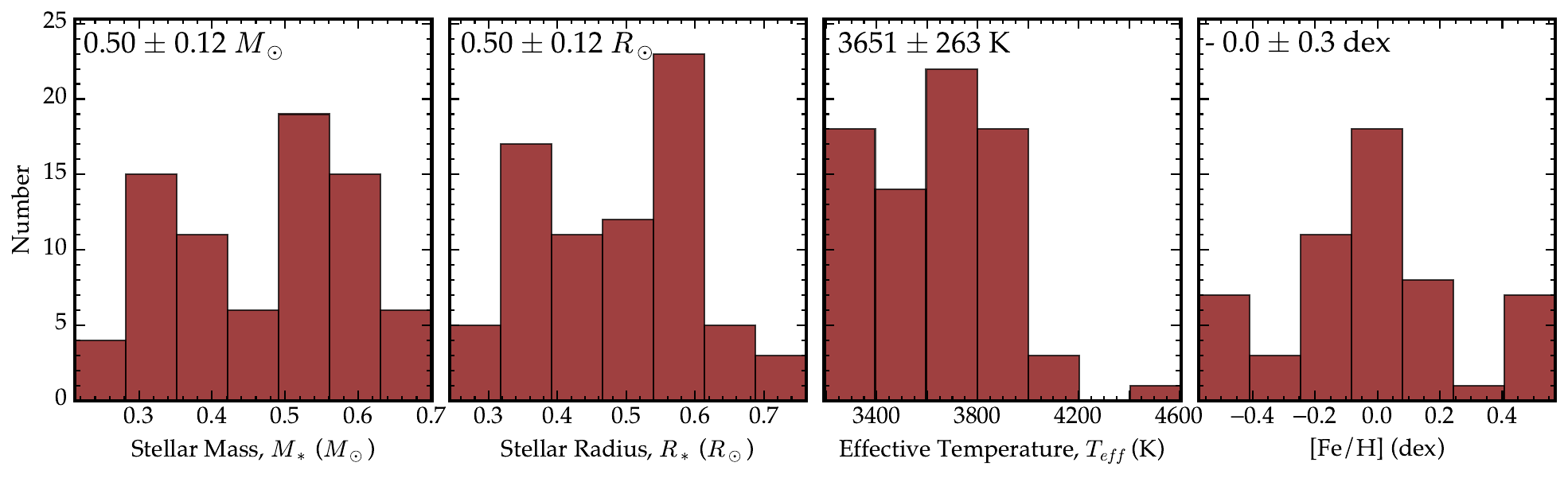}
    \caption{Histograms of measured stellar parameters for the \ntois{} stars in our sample, showing (from left to right) stellar masses, radii, effective temperatures, and metallicities. The median values and associated standard deviations are shown in each panel.}
    \label{fig:stellarparams}
\end{figure*}

Finally, we calculated the stellar effective temperature using the Stefan-Boltzmann law. The stars in our sample have masses between 0.21 and 0.70 $\rm{M_\odot}$, with a median stellar mass of 0.50 $M_\odot$ and standard deviation of 0.12 $M_\odot$. The measured effective temperatures range between 3190 and 4600 K, with a median effective temperature of 3650 K and a standard deviation of 260 K, corresponding to early M-type dwarf stars. We report our updated stellar parameters in Table \ref{tab:updated_params}, and show the stellar parameter ranges included in our sample in Fig. \ref{fig:stellarparams}.

\section{Planets}\label{sec:planets}
\subsection{Planet Parameters} \label{sec:updatedplparams}

There are \nPCs{} as-yet unconfirmed planet candidates (TESS Objects of Interest, TOIs) orbiting 23 stars in our sample. TOIs typically have large radius uncertainties reported on ExoFOP, as the TESS project pipeline takes a generic approach to identifying planet candidates. Therefore, we independently fit the transit data of these TOIs in order to improve the radius measurements. We used the \texttt{lightkurve} package \citep{Lightkurve_2018} to download the 2-minute SPOC TESS light curves for each of these TOIs. We then normalized and stitched together the light curves from each TESS Sector. For five of the 23 planet candidate hosting TOIs, the TESS mission has collected over 20 sectors of data. 

To constrain the size and orbital properties of the planet candidates, we modeled the transit photometry using the \texttt{exoplanet} package \citep{exoplanet:joss}. We defined our transit model using the planet-to-star radius ratio ($\rm{R_p/R_*}$), transit epoch ($\rm{T_0}$), orbital period ($\rm{P}$), semi-major axis in units of stellar radii ($\rm{a/R_*}$), impact parameter (b), and quadratic limb-darkening coefficients \citep[$\rm{q_1}$, $\rm{q_2}$, using the parameterization of][]{kipping+2013MNRAS}. Our model also included a mean baseline flux term ($\langle F \rangle$). Finally, we assumed a circular orbit for the planets, setting the eccentricity equal to zero.

\begin{table}
    \centering
    \caption{Prior distributions used to fit transits}
    \begin{tabular}{lc}
    \hline
    \hline
        \textbf{Parameter} & \textbf{Prior Distribution} \\
        \hline
        ln(Orbital Period), ln(P) & $\mathcal{N}(\mu, \sigma = 0.01)$ \\
        Transit Epoch, $\mathrm{t_0}$ &  $\mathcal{N}(\mu,  \sigma = 0.1))$ \\
        Mean flux, $\langle \mathrm{F} \rangle$ &  $\mathcal{N}(\mu = 1.00,  \sigma = 0.01)$\\
        Limb darkening coeff., $q_{1}$ & $\mathcal{U}(0, 1)$ \\
        Limb darkening coeff., $q_{2}$ & $\mathcal{U}(0, 1)$ \\
        Planet-star radius ratio, $\mathrm{R_p/R_*}$ & $\mathcal{U}(0.01, 0.3)$ \\
        Impact parameter, b & $\mathcal{U}(0, 1+\mathrm{R_p/R_*})$ \\
        \hline
    \end{tabular}
    \label{tab:plfit_priors}
\end{table}

We optimized the model parameters using Bayesian inference, implementing a Hamiltonian Monte Carlo (HMC) No U-Turn Sampler \citep[NUTS,][]{Hoffman+Gelman_2011} with \texttt{PyMC3} \citep{exoplanet:pymc3} to sample the posterior probability distributions. The prior distributions used for each parameter are given in Table \ref{tab:plfit_priors}, and we used the reported ExoFOP planet parameters as test values for each parameter. We set the target acceptance rate to 0.95 (to account for the higher acceptance fractions returned by HMC samplers compared with Metropolis-Hastings samplers) and initialized the sampler by adapting a dense mass matrix from the sample covariances. We then ran the sampler using 2 chains, each drawing 10,000 samples after discarding 2,000 burn-in steps. To check for convergence, we computed the Gelman-Rubin Diagnostic and visually inspected the sampler trace plot for each parameter. The median and 68\% confidence range for each parameter are given in Table \ref{tab:pcfits}. 

\begin{figure}
    \includegraphics[width=\columnwidth]{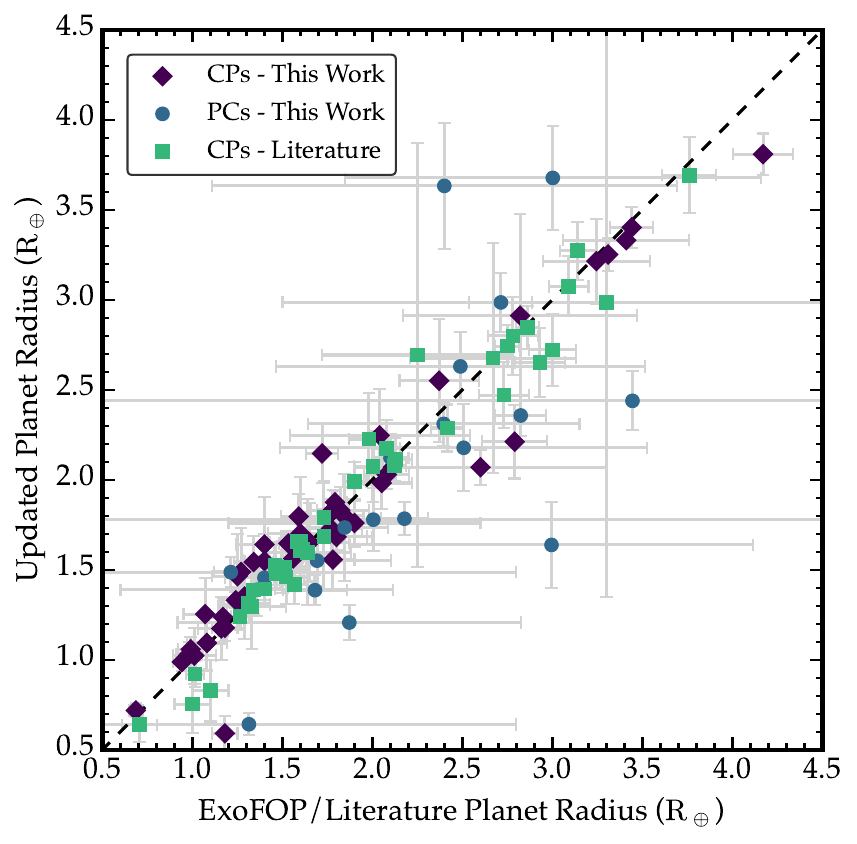}
    \caption{Comparison of existing and updated planet radii for small ($\rm{R_p} < 4 R_\oplus$) planets orbiting M dwarfs. The shape of data points is set by whether the planets' host stars are in the sample presented in this work (dark purple diamonds - confirmed planets, blue circles - candidate planets) or with existing stellar parameters measured using IRTF/SpeX (green squares). A one-to-one line (dashed black) is included for reference.}
    \label{fig:radiuscomp}
\end{figure}

We used the results of our transit fits (described above) and the measured stellar radii (described in Section \ref{sec:methods_phot}) to calculate updated planet radii. We also recalculated the radii of confirmed planets in our sample using the stellar radii reported in this work and literature planet-to-star radius ratios. We report the orbital periods, planet radii, impact parameters, planet-to-star radius ratios and times of conjunction in Table \ref{tab:pcfits} for planet candidates and Table \ref{tab:cpfits} for confirmed planets. In Fig. \ref{fig:radiuscomp}, we compare our new radius measurements with existing values from ExoFOP for planet candidates and from the papers referenced in Table \ref{tab:cpfits} for confirmed planets. For the \nPCs{} planet candidates, we improved the radius measurement precision for 19 planets, with a median precision improvement of $\sim80\%$. This improvement is largely driven by the increased precision on the measured transit depths from our fits compared to the transit depths reported on ExoFOP.

\subsection{Planet Vetting} \label{sec:validation}

Of the \nCPs{} confirmed planets in our sample, three have been validated using colour validation \citep{Esparza-Borges2022b, Pelaez-Torres2024}. This method depends on the fact that planetary transits are achromatic, while apparent transits due to other astrophysical scenarios (e.g. a blended eclipsing binary) are wavelength-dependent \citep{Rosenblatt1971, Tingley2004}. Therefore, multi-band photometry of transit events can determine whether they are caused by planets. A further 36 planets have been validated using statistical validation \cite[e.g.][]{CastroGonzalez2020}, which compares the observed transit to simulated transits caused by non-planetary phenomena. The remaining planets have been confirmed using radial velocity observations to obtain mass measurements.

For the \nPCs{} planet candidates in our sample, we used the \texttt{triceratops} \citep{giacalone+21} statistical validation Python package to characterize these candidates. This method calculates the probabilities of a range of transit-producing scenarios using the observed  transit alongside imaging and photometric observations of the host star to characterize nearby contaminating stars. For this analysis, we folded each light curve to the orbital period and transit epoch of the associated TOI, and binned the light curve by a factor of 1200. We also only used the region of the light curve within $\pm \mathrm{P}/8$ days of the transit (where P refers to the planet's orbital period), for computational efficiency. We included the imaging observations described in Section \ref{sec:appendix_imaging} to identify close stellar companions to the target stars, which would have a diluting effect on the observed transits. We are not able to conclusively validate any of the planet candidates in our sample, but flag five as likely planets (TOIs 233.01, 1638.01, 4325.01, 5511.01, and 5961.01). To do this, we followed the criteria set out in \citet{giacalone+21}, namely that likely planets have false positive probabilities (FPP) $<$ 0.5 and nearby false positive probabilities (NFPP) $< 10^{-3}$. We detail these results in Table \ref{tab:trivetting}. We note that the probabilities for each target do not sum to one, as we report the median probabilities after running \texttt{triceratops} 20 times on each target.

\begin{deluxetable*}{lcccccccccc} 
\tabletypesize{\scriptsize}
\tablecaption{Scenario Probabilities for Likely Planets}
\tablehead{
\colhead{TIC}&\colhead{TOI}&\colhead{P (days)} &\colhead{$\mathrm{t_0}$ (BJD-2457000)}&\colhead{Transit Depth (ppm)}&\colhead{$P_{TP}$}&\colhead{$\sigma_{P_{TP}}$}&\colhead{$P_{FPP}$}&\colhead{$\sigma_{P_{FPP}}$} &\colhead{$P_{NFPP}$}&\colhead{$\sigma_{P_{NFPP}}$} \label{tab:trivetting}}
\startdata
415969908 & 233.01 & 11.7 & 1365.26 & 2780 & 0.88  & 0.01  & 0.24  & 0.02 & 0.00 & 0.00 \\	
202185707 & 4325.01 & 1.65 & 1469.37 & 2710 & 0.87 & 0.02 & 0.26 & 0.04 & 0.00 & 0.00 \\
262689575 & 5961.01 & 1.62 & 2447.79 & 342 & 0.96 & 0.02 & 0.07 & 0.05 & 0.00 & 0.00 \\
312862941 & 1638.01 & 0.92 & 2908.44 & 4305 & 0.87 & 0.03 & 0.18 & 0.03 & 0.00 & 0.00 \\
389040826 & 5511.01 & 4.72 & 2575.04 & 11520 & 0.94 & 0.03 & 0.13 & 0.06 & 0.00 & 0.00 
\enddata
\end{deluxetable*}

\section{Discussion} \label{sec:discussion}

\subsection{Comparison with Literature Stellar Parameters} \label{sec:lit}

\begin{figure}
    \includegraphics[trim={21cm 0 0 0},clip, width=\columnwidth]{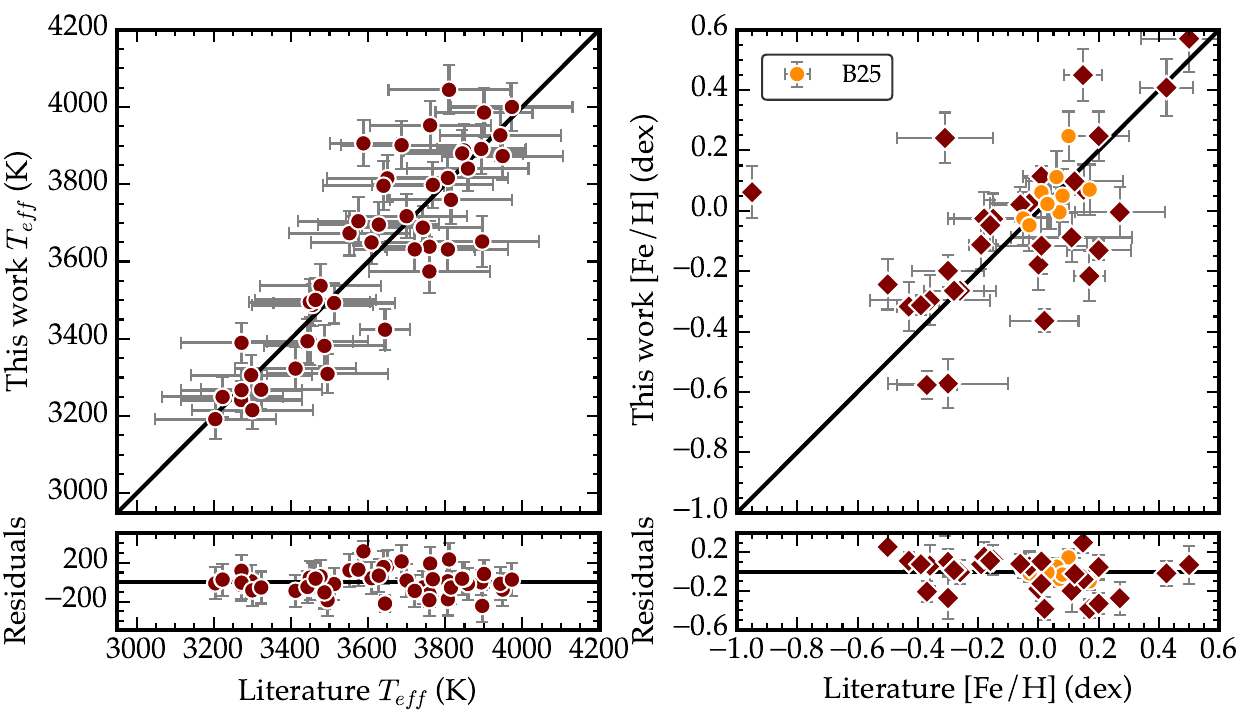}
    \caption{Top: Comparison of literature values from the TIC and measured in this work for stellar effective temperature and metallicity ([Fe/H]). We include a 1:1 line (black) for reference, and highlight stellar metallicity measurements from \citet[][B25]{Behmard2025} in orange in the right-most plot. The effective temperatures derived from our spectra have markedly smaller errors than the corresponding literature values. Our measured metallicities are in moderate agreement with literature values, with some exceptions discussed in Section \ref{sec:lit}. Bottom: Residuals of our derived values from the catalog measurements. The outliers discussed in Section \ref{sec:lit} are excluded from the residual plots for visual clarity.}
    \label{fig:stellarcomp}
\end{figure}

Measuring M dwarf stellar properties is notoriously challenging. Their inherent faintness and high activity inhibit photometric characterization, while complex molecular features due to their cool atmospheres hinder spectroscopic studies. Measurements of fundamental stellar properties, namely mass, radius, effective temperature and metallicity, are subject to both intrinsic scatter and systematic errors.

We uniformly measured fundamental stellar parameters of the stars in our sample using IRTF/SpeX spectra, as detailed in Section \ref{sec:methods} and reported in Table \ref{tab:updated_params}. Our sample is made up of low-mass stars, primarily M dwarfs, with a median stellar mass of 0.50 $\pm 0.12 M_\odot$ and median effective temperature of 3650 $\pm$ 260 K. We measure a range of iron abundances ([Fe/H]) between -0.57 and +0.57 dex, and of total metallicity ([M/H]) between -0.48 and +0.42 dex.

In Figure \ref{fig:stellarcomp}, we compare our measured stellar parameters with those from the TIC \citep{Paegert2021}. All of the targets in our sample are included in the Cool Dwarf Catalog \citep[CDC,][]{muirhead+18, Chittidi2019}, which was incorporated into the the TIC Candidate Target List (CTL) by \citet{Stassun2018, Stassun2019}.

Fundamental stellar parameters were measured for targets in the CDC via the following methods, as described in \citet{Stassun2019}. Stellar mass was calculated following the mass-magnitude relationship presented in \citet{Mann19}. This is the same relation that we used in Section \ref{sec:methods_phot} to calculate the masses of our targets. Therefore, the residuals between the literature stellar masses and those reported in this work are, as expected, negligible, and are all consistent with 0 to 1 $\sigma$. Similarly, stellar radii were measured for CDC targets using the radius-magnitude relation presented in \citet{mann+15}, as we did for the targets presented in this work. However, the CDC stellar radius calculation does not include the [Fe/H] correction that we incorporated. \citet{mann+15} report that the inclusion of a multiplicative metallicity correction resulted in a small but statistically significant improvement in the resulting radius-magnitude fit. 

The effective temperatures of stars in the CDC were calculated using $\rm{G_{BP}}$ and $\rm{G_{RP}}$ magnitudes and custom photometric relations based on those reported in \citet{mann+13d}. This relation was developed to calculate effective temperature for stars without measured parallaxes. However, our targets are nearby cool stars with measured distances, and we calculated effective temperature via the Stefan-Boltzmann law (see Section \ref{sec:methods_phot}). We find very good agreement between the literature effective temperatures and those measured in this work. The effective temperature residuals are consistent with 0 for the majority of targets (38/\ntargets{}). We also improve the effective temperature precision by a factor of $\sim2.5$, with the median uncertainty decreasing from 158 K to 59 K.

We also compared our measured stellar metallicities with literature values, which were available for 23 of our targets. \citet[][henceforth B25]{Behmard2025} presented a large catalog of M dwarf metallicities derived using an application of The Cannon \citep{cannon, canoonv2}. This is a data-driven approach of measuring metallicities by training a machine learning model on high-resolution spectra from the Sloan Digital Sky Survey-V/Milky Way Mapper \citep[SDSS-V/MWM,][]{Gunn2006, Wilson2019, Kollmeier2026}, with spectra from the Apache Point Observatory Galactic Evolution Experiment \citep[APOGEE,][]{Majewski2017}. The metallicity measurements reported in B25 have an associated uncertainty of 0.02 dex. Ten of the targets in our sample were included in this catalog, and we compared the resulting metallicities in Fig. \ref{fig:stellarcomp}, finding that the [Fe/H] values are in moderate agreement, with a few notable outliers. We attribute the moderate differences between our measured metallicities and available literature values as being due to the methods used to calculate [Fe/H]. The majority of literature metallicities were calculated using ExoFASTv2 \citep{exofastv2} to model spectral energy distributions of target stars. This approach uses sparser information than our approach, and therefore returns less reliable metallicity measurements.

K2-344 (TIC 203289099, EPIC 212081533) has a literature stellar metallicity of [Fe/H] = -0.95 $\pm$ 0.02 dex \citep{deLeon2021}, which is highly discrepant from our measured value of [Fe/H] = 0.06 $\pm$ 0.09 dex. We attribute this large difference to the use of the kea grid of synthetic stellar spectra \citep{kea} in \citet{deLeon2021} to provide spectroscopic constraints on stellar metallicity. The kea models are calibrated to stars with effective temperatures between 5000 and 6700 K, while K2-244 is a cool dwarf with an effective temperature of 3816 $\pm$ 63 K (this work). Therefore, the stellar parameters returned using the kea models are likely inaccurate for this target. 

K2-155 (TIC 17307715, EPIC 210897587, LP 415-17, TOI-5562) has a literature stellar metallicity of -0.3 $\pm$ 0.2 dex \citep{DiezAlonso2018a} and mass of $0.65^{+0.06}_{-0.03} \rm{M_\odot}$, measured using HARPS-N spectra \citep{Cosentino2012}. These values are consistent with those reported in \citet{Hirano2018b} ([Fe/H] = -0.42 $\pm$ 0.12 dex, $M_* = 0.540 \pm 0.056 M_\odot$) using the Tull Coud\'e Spectrograph \citep{Tull1995} on the McDonald Observatory 2.7m Harlan J. Smith Telescope. In this work, however, we measured a stellar metallicity of -0.57 $\pm$ 0.08 dex, which is $\sim 1.5 \sigma$ discrepant from the literature values. This difference may be due to the wavelength coverage of the instruments used: HARPS-N and the Tull Coud\'e Spectrograph are primarily optical instruments compared to the SpeX NIR coverage. Previous work \citep[e.g.][]{Rajpurohit2018, Johnson2009} highlighted possible systematic offsets between optical and IR metallicities, due to complex M dwarf spectra and methodological challenges in obtaining high SNR spectra of a given target in both wavelength regimes. 

\begin{figure*}
    \includegraphics[width=\textwidth]{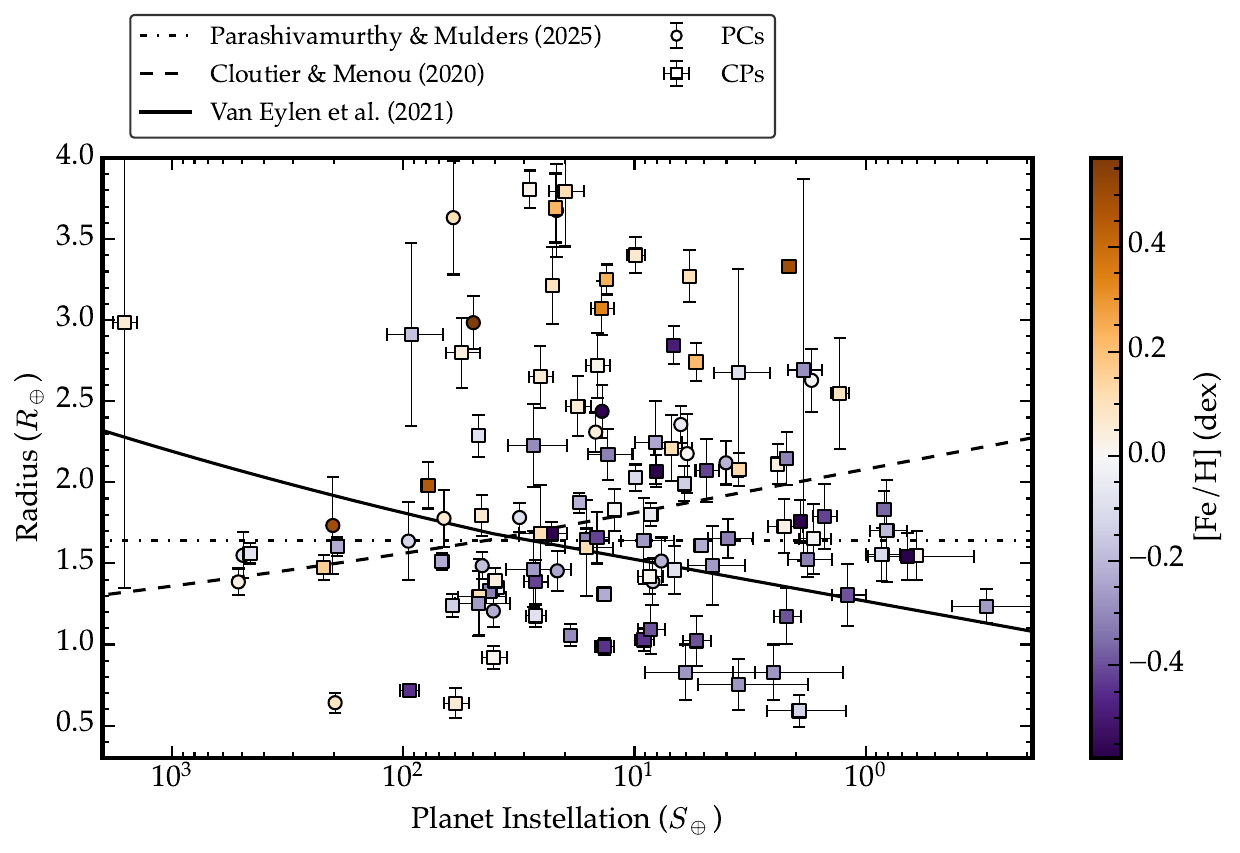}
    \caption{Planet instellation vs. radius for the observed planet candidates (circles) and confirmed planets (squares) in our sample, as well as confirmed planets from the literature (squares). Points are coloured according to measured stellar metallicity [Fe/H]. Planets orbiting metal-rich stars tend to be larger than those orbiting metal-poor stars. Empirical relations for the radius valley around low-mass stars are shown by solid \citep{vaneylen21}, dashed \citep{Cloutier2020}, and dash-dotted \citep{Parashivamurthy2025} lines, based on completeness-corrected samples of M dwarfs We note that several of the confirmed planets have substantial unceranties in planet radius exceeding $20\%$. These uncertainties in are due to large measurement errors in the transit depth reported in the liteature (see references in Table \ref{tab:cpfits}) and not from the stellar radius uncertainties presented in this study, which are nominal ($~4\%$).}
    \label{fig:pop}
\end{figure*}

\subsection{Stellar Metallicity -- Planet Radius for Small Planets ($R_p < 4 R_\oplus$)}

In order to expand our sample, we identified 28 red dwarfs hosting 43 planets and planet candidates with metallicities measured using IRTF/SpeX spectra and the same methodology as in this paper. These results were published in \citet{dressing+19} and \citet{Gore2024}. By constructing a sample in this way, we avoided introducing systematic errors in metallicities measured using different instruments or techniques. The full sample contained a total of \nplanetsottal{} planets and planet candidates orbiting \nStartotal{} cool dwarf stars. Of these, 108 planets and planet candidates orbiting 72 cool dwarfs had radii $< 4 \rm{R_\oplus}$. In Fig. \ref{fig:pop}, we plotted planet radius as a function of planet instellation and stellar metallicity, showing that cool dwarfs in our sample which host larger planets tend to be metal-rich. The sample also contained seven giant planets, with radii $> 9 \rm{R_\oplus}$. As described in Section \ref{sec:updatedplparams}, we uniformly re-derived planet radii using stellar radii measured using IRTF/SpeX observations and empirical photometric relations. The stars in the expanded sample span the same stellar mass, radius, and effective temperature ranges as those presented for the first time in this work (see Section \ref{sec:methods}). We use this expanded sample for the remainder of the discussion.

Unlike the well-known correlation between giant planet occurrence and stellar metallicity \citep{fischer+valenti05}, the relationship between small planet occurrence and stellar metallicity has not been conclusively settled. Characterizing this relationship is important both for understanding how small planets form, but also to enable conditional occurrence rates which probe how gas giant formation is impacted by the presence or absence small planets in the system \citep{Bryan2025}.

Theoretical results suggest that planet formation is broadly suppressed in metal-poor environments \citep[e.g.][]{mordasini12, burn2021, xenos25}. \citet{Wang2015a} reported an enhanced occurrence rate for both super-Earths and sub-Neptunes around metal-rich solar-type stars compared to a metal-poor stellar sample. Additionally, \citet{Boley2024} report evidence of a sharp decrease in small (1 - 3 $\rm{R_\oplus}$) planet occurrence for FGK stars with metallicities lower than [Fe/H] = -0.31 $\pm$ 0.02 dex.  When considering only cool dwarfs, \citet{Lu2020} found a linear relationship between small planet occurrence and stellar metallicity around low-mass stars. They reported that for planets with radii between 2 and 5 $\rm{R_\oplus}$, planet occurrence scales linearly with stellar metallicity. They also predicted that planets $<2 \rm{R_\oplus}$ should be very rare around early-type M dwarfs with [M/H] $<$ -0.5 dex. Conversely, \citet{Kutra2021} find that the occurrence of small Kepler planets (1 - 4 $R_\oplus$) is independent of stellar metallicity. 

Our sample is consistent with the results suggesting a decreased planet occurrence around the most metal-poor stars, although we do not correct for observational biases in our sample and so do not calculate occurrence rates.  Our sample does not include any planet hosting stars with [M/H] $<$ -0.5 dex, and contains only 2 targets with [Fe/H] $<$ -0.31 dex.  This is in line with the broad picture of planet formation, which requires a sufficient solid mass budget to form planets at all \citep{lee2015}. However, while a measurement of the occurrence of small planets around M dwarfs as a function of stellar metallicity is beyond the scope of this work, we perform an intra-sample comparison of stellar metallicity distributions, described below.

\subsection{Sub-Neptune vs. Super-Earth Host Metallicities} \label{sec:discussion_perplanet}

The cool dwarfs in our sample mostly host planets with radii $< 4 \rm{R_\oplus}$, which can further be split into sub-Neptunes and super-Earths. It remains unclear whether super-Earths and sub-Neptunes are formed as distinct populations, and whether these planets form differently around low-mass stars than around solar-type host stars \citep{Parc2024}. Around low-mass stars, super-Earths may primarily be rocky planets that formed in situ (i.e., close in to their host stars where volatiles in the protoplanetary disk are vaporized). In this picture, sub-Neptunes around low-mass stars may be volatile-enriched ``water worlds'' that formed beyond the ice line and subsequently migrated inwards \citep{Venturini2020, Schlecker2021, Alessi2020c, Luque2022, Burn2024}. If the separation between super-Earths and sub-Neptunes is compositional, as evidenced by the empirical density valley around M dwarfs but not FGK stars \citep{Luque2022, Parc2024}, this may be reflected in their host stars' metallicities. We therefore compare the stellar metallicity distributions between the super-Earth and sub-Neptune hosts in our sample. 

In order to split the planet sample into super-Earths and sub-Neptunes, we considered six definitions of the radius valley, three empirical relations and three theoretical relations. We considered a radius cut-off of 1.64 $\rm{R_\oplus}$ as reported in \citet{Parashivamurthy2025} (henceforth PM25), the negative-slope radius-instellation relation of \citep{Cloutier2020} (henceforth CM20), and the positive-slope radius-instellation relation of \citep{vaneylen21} (henceforth V21). These relations are all empirically motivated for low-mass stars. PM25 derived their radius valley relation using 843 M dwarfs observed by TESS, with a median stellar mass of 0.65 $M_\odot$. CM20 studied 17,393 early M dwarfs observed by Kepler and K2, with a median stellar mass of 0.67 $M_\odot$, and V21 considered 27 M dwarfs with well-characterized planets and a median stellar mass of 0.33 $M_\odot$. The PM25 and C20 samples are well-matched with ours, which has a median stellar mass of 0.50$\pm$ 0.12 $\rm{M_\odot}$. We note that the radius valley is not visible in our sample (see Fig. \ref{fig:pop}); this is due to the lack of correction for observational biases in our sample and relatively small planet sample. We constructed our sample by selecting stars with known planet and/or planet candidates. As such, we emphasize that this work does not encompass the underlying planet population or seek to compare the metallicity of planet-hosting stars to those without planets. Rather, we compare the intra-sample metallicity distributions of observed super-Earth and sub-Neptune hosts.

\begin{figure*}  
    \includegraphics[width=\textwidth]{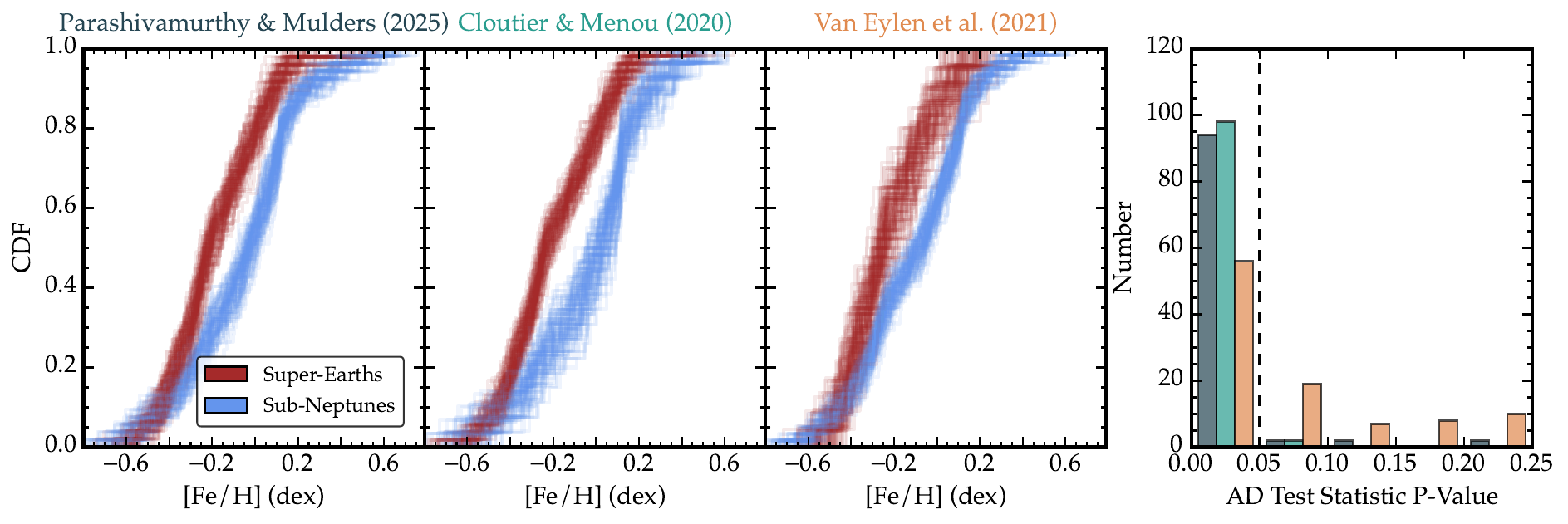}
    \includegraphics[width=\textwidth]{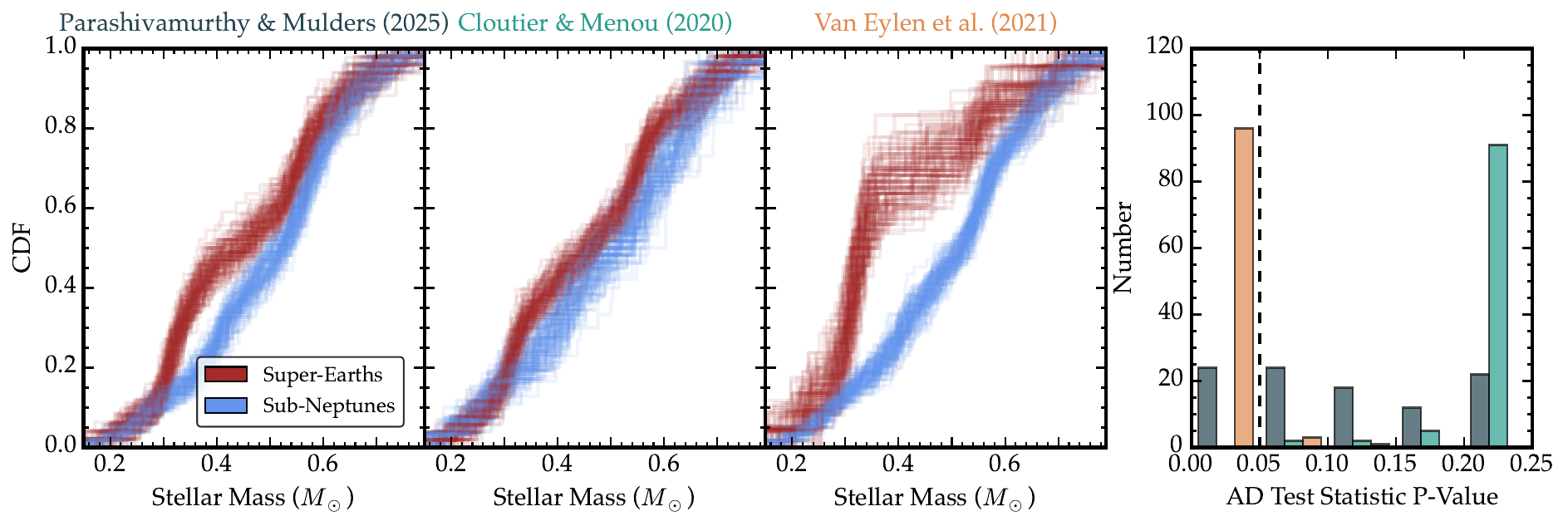}
    \caption{(Left 3 plots:) Per planet cumulative distribution function of measured metallicities (top row) and stellar mass (bottom row) for super-Earths (brown) and sub-Neptunes (blue) for 100 samples of the measured planet radius, stellar metallicity, and stellar mass uncertainties. The distinction between the two populations are set following the empirical radius valley prescriptions from PM25 (dark blue),  CM20 (teal), and V21 (orange). (Right:) Histograms showing the p-values associated with the k-sample Anderson-Darling  \citep[AD,][]{ksamp} tests for the stellar metallicity (top) and mass (bottom) distributions for each of 100 draws from the planet sample plotted in Fig. \ref{fig:pop}. A vertical dashed black line at a p-value of 0.05 indicates a typical upper threshold used to reject the null hypothesis under an AD test. These p-value distributions show we can reject the null hypothesis that the stellar metallicity distributions are drawn from the same underlying population with high confidence for the all of the radius valley relations. Additionally, we cannot reject the null hypothesis that the samples are likely drawn from the same stellar mass population for the CM20 radius valley prescription.}
    \label{fig:cdf}
\end{figure*}

In Fig. \ref{fig:cdf}, we plot the cumulative distribution functions (CDFs) of the stellar metallicities in each sub-sample. To account for the planet radius and stellar metallicity measurement uncertainties, we calculated the CDFs  for the super-Earth and sub-Neptune populations 100 times, sampling the planet radius and [Fe/H] measurements for each planet and corresponding host star. We performed this exercise using the three empirical radius valley scalings discussed above, as well as three theoretical radius valley scalings (see Appendix \ref{sec:appendix_theory}). Given that the theoretical scalings do not incorporate a stellar mass dependence, we primarily discuss the results for the empirical radius valley scalings.

The CDFs visually show that the sub-Neptune hosts are typically more metal-rich than the super-Earth hosts for all of the radius valley prescriptions considered. To characterize this observed directionality, we compute the signed difference,
\begin{equation}
\Delta F(x) = F_{\rm SES}(x) - F_{\rm SNS}(x),
\end{equation}
where \(F_{\rm SES}(x)\) and \(F_{\rm SNS}(x)\) are the CDFs evaluated at metallicity \(x\). Positive values of \(\Delta F(x)\) indicate that the super-Earth sample contains a larger fraction of objects with \([\mathrm{Fe/H}] \le x\), implying a shift toward lower metallicities relative to the sub-Neptune hosts, while negative values indicate the opposite. We summarize the overall directional offset by integrating the CDF difference, providing a measure of the overall shift in metallicity between the two samples. We find that for all radius valley prescriptions considered, the integrated signed CDF difference is $>0$, demonstrating that the sub-Neptune hosts are typically more metal-rich than the super-Earth hosts. This trend remains when we resampled the population of planets via jack-knife cross-validation, iteratively selecting 80$\%$ of the our sample. In order to statistically confirm this distinction in stellar metallicity distributions, we calculated the k-sample Anderson-Darling (AD) test statistic \citep{ksamp} for each draw of the sub-Neptune and super-Earth populations. This statistic tests the null hypothesis that metallicity samples are drawn from the same population without needing to specify the form of the underlying distribution. We show the histogram of the p-values associated with these tests in Fig. \ref{fig:cdf}. For the theoretical scalings, we find that the null hypothesis can be rejected more confidently under the gas-depleted formation scenario presented in \citet{Lopez2018}, and can be rejected less frequently for the photo-evaporative mass loss \citep{Lopez2018} and core-powered mass loss \citep{Gupta2019} scalings (see Fig. \ref{fig:theorycdf}). 

We also investigate whether the observed metallicity enhancement is due to other aspects of the stellar sample. Specifically, we use the AD test statistic to compare the stellar mass, effective temperature, and luminosity distributions of the host stars. We include the corresponding stellar mass CDF plots in Fig. \ref{fig:cdf} as a representative example. The resulting p-value distributions show that the bulk of the p-values under the CM20 case are $>0.05$, and that we can therefore not reject the null hypothesis that the two stellar mass samples are drawn from the same underlying distribution. We interpret this result as suggesting that the observed metallicity results in the CM20 case are unlikely to be due to differences in the stellar mass, and there is some evidence that this is also the case for the PM25 radius valley definition. However, under the V21 radius valley prescription, we can reject the null hypothesis that the stellar sub-samples are drawn from the same underlying population, with the p-value distribution strongly peaked $<0.05$. This is likely an observational bias, as the V21 radius valley was constructed using a sample of planets orbiting M dwarfs with well-characterized masses and radii. Super-Earths are typically less massive than sub-Neptunes, and thus are preferentially detected using radial velocities around lower-mass stars, as seen in the stellar mass CDF in the V21 case. Therefore, the observed discrepancy in stellar metallicities in the V21 case is not a robust result. We find the same results for the effective temperature and luminosity CDFs.

In this work, we observe that sub-Neptunes preferentially orbit metal-rich low-mass stars relative to super-Earths. Sub-Neptune formation and interior structure remain open questions. Given the rapid inward drift and accretion of solid material in the inner regions of protoplanetary disks \citep{ogihara+15}, migration is highly likely to have played a role in shaping the observed population. However, it is unclear where the solid building blocks of sub-Neptunes are primarily assembled \citep[e.g.][]{Johansen2017, Bean2021}: solid material may migrate inwards before aggregating into planets (i.e., drift model), or planetary cores may form further out in the disk before experiencing inward migration (i.e., migration model). If sub-Neptunes orbiting M dwarfs are water-worlds, their volatile-rich interiors are most consistent with the migration model, as pebbles migrating inwards would lose their volatile contents prior to assembly under the drift model \citep{Bean2021, bitsch2019, Venturini2020, Venturini2024}.

We note that other mechanisms have been invoked to explain the radius valley around FGK stars, including photo-evaporative mass loss \citep{Owen2013}, core-powered mass loss \citep{Gupta2019}, gas-depleted formation \citep{Lopez2018}, and gas-poor formation \citep{Lee2022}. These processes are also able to reproduce the radius valley around FGK stars, and may also lead to distinct stellar metallicity distributions between super-Earth and sub-Neptune host stars. Both photo-evaporative and core-powered mass loss are suppressed in high metallicity atmospheres \citep{owenmurrayclay2018, Gupta2020}. If metal-rich stars host planets with higher atmospheric metallicities which thus experience less mass loss, this would also produce the enhanced sub-Neptune occurrence around metal-rich FGK stars \citep{petigura+2018}, and the observed metal-rich sub-Neptune host M dwarfs presented in this work. The gas-depleted and gas-poor formation scenarios also could reproduce this result, as solid cores are able to form more rapidly and thus accrete gas to form sub-Neptunes around metal-rich stars with enhanced metal content in their disks. However, in all of these scenarios, the sub-Neptunes produced are likely gas dwarfs, with solid cores surrounded by H/He envelopes. This is in conflict with the observed density valley for planets around M dwarfs, which strongly suggests a compositional difference between super-Earths and sub-Neptunes orbiting low-mass stars. Additionally, \citet{cherubim2023} found that the compositions of seven close-in small planets orbiting M dwarfs disfavor thermally-driven mass-loss mechanisms. Therefore, we consider these mechanisms as less likely to have formed the small planets in our sample, although distinct stellar metallicity distributions alone are not sufficient to discriminate between different formation pathways, and we do not present an occurrence result in this work.

In order to connect the observational trend presented in this work with underlying theories of planet formation, we also consider results from planet population synthesis models. The Next-Generation Planet Population Synthesis model \citep{emsenhuber21} incorporates a diverse set of physical processes, including planetesimal accretion and migration through the disk. \citet{burn2021} found that planets formed around low-mass stars under this model also display a bimodal density distribution made up of dry, rocky super-Earths and wet sub-Neptunes. They also found that the mean metallicity of low-mass ($M_* < 1 \rm{M_\odot}$) stars hosting sub-Neptunes was higher than that of stars hosting super-Earths. Higher metallicities may be required to form sub-Neptunes around low-mass stars due to the relatively small disk masses and thus small solid reservoirs available for planet formation. 

\citet{Bitsch2016} found that water-rich planet formation is enhanced in disks with low water-to-silicate ratios. They find that disks with large refractory reservoirs have high Rosseland mean opacities across the disk, unlike water-dominated disks where the opacity decreases inside of the ice line as icy grains sublimate. The increased opacity in low water-to-silicate disks inhibits gas accretion onto protoplanets and thus their subsequent outward migration. This suggests that both the refractory and water content beyond the ice line is important in forming water-rich planets. Therefore, sub-Neptune formation via the accretion of water-rich pebbles beyond the snow line would be enhanced in metal-rich disks. If super-Earth and sub-Neptune cores were primarily made up of refractories (i.e., with limited or no water), we would not expect there to be evidence of distinct host star metallicity distributions. The enhanced metal content of sub-Neptune-hosting M dwarfs presented in this work implies that additional material (namely, water) is involved in sub-Neptune formation around M dwarfs. Future, sensitivity-corrected studies will enable a demographic study of this topic, teasing apart the occurrences of both super-Earths and sub-Neptunes orbiting M dwarfs as a function of stellar metallicity. 

\subsection{Sub-Neptune vs. Super-Earth-Only System Metallicities}

In the previous section, we compare the metallicities of M dwarfs that host super-Earths and sub-Neptunes. There are, of course, M dwarfs that host both super-Earths and sub-Neptunes, or multiple of either type of planet. In our sample there are 19 multi-planet systems, 4 of that host more than 3 planets. These stars which host multiple transiting planets are thus counted repeatedly in the CDFs calculated and shown in Fig. \ref{fig:cdf}. Above, we discuss a formation scenario wherein sub-Neptunes around M dwarfs form in metal-rich disks beyond the ice line prior to inward migration. In this picture, we expect to see a discrepancy between the metallicity distributions of stars which host only super-Earths and stars which host any number of sub-Neptunes, as systems that did not have some critical metal inventory could not have formed any sub-Neptunes. 

\begin{figure*}  
    \includegraphics[width=\textwidth]{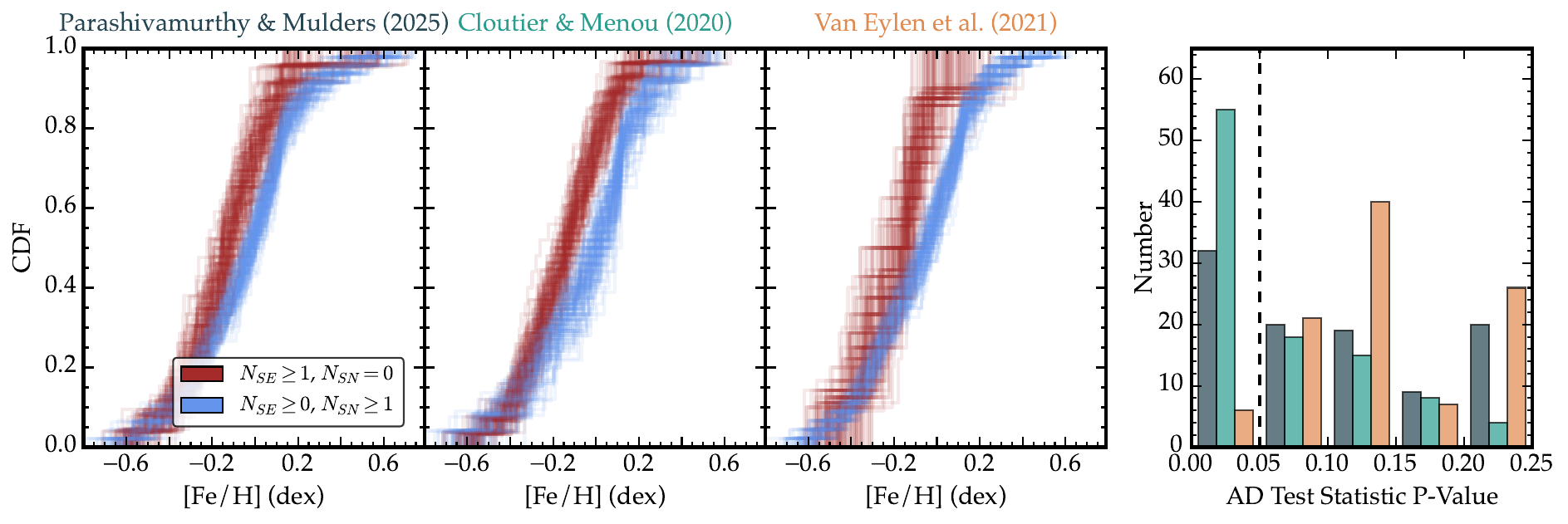}
    \includegraphics[width=\textwidth]{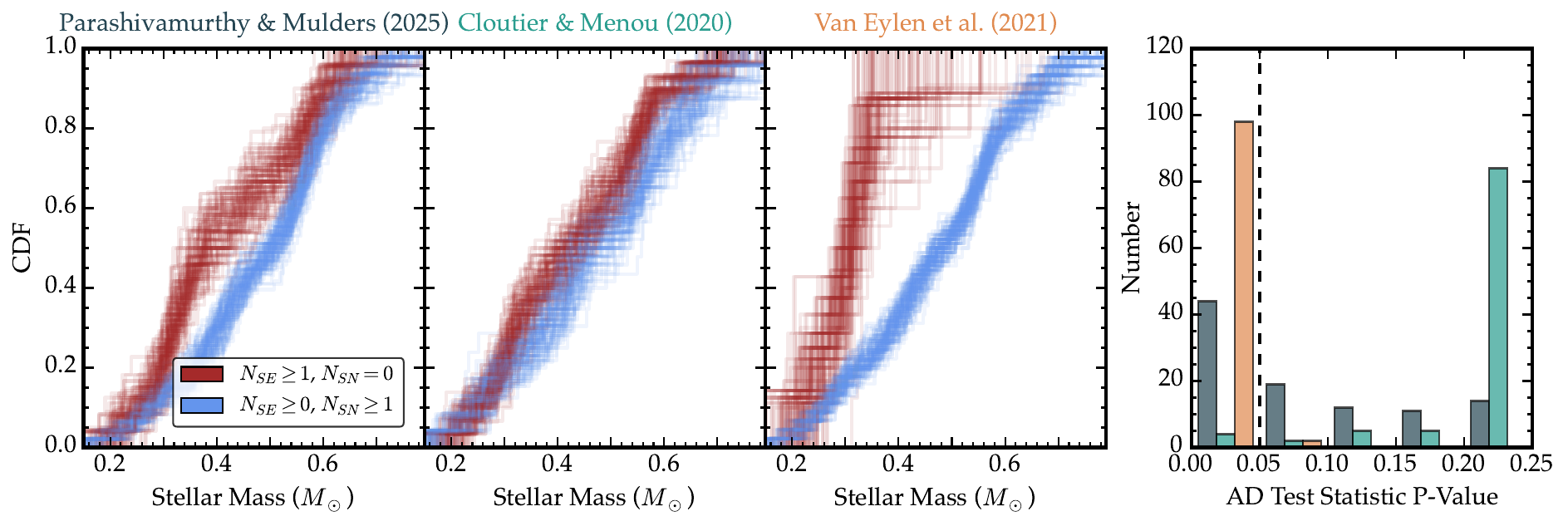}
    \caption{CDFs of stellar metallicity and mass, as in Fig. \ref{fig:cdf}, but plotted on a per system basis rather than a per planet basis. Super-Earth-only systems are shown in brown and any-sub-Neptune systems are shown in blue. The p-value distributions show that we can reject the null hypothesis, namely that the per system stellar metallicity distributions are drawn from the same underlying population, for the PM25 and CM20 radius valley relations. Additionally, we cannot reject the null hypothesis that the stellar mass samples are drawn from the same distribution under the CM20 radius valley prescription.}
    \label{fig:cdf_persystem}
\end{figure*}

We plot the CDFs of stellar metallicity on a per system basis in Fig. \ref{fig:cdf_persystem} as opposed to the per planet basis depicted in Fig. \ref{fig:cdf}. In this case, we consider the metallicity distribution of stars hosting at least one super-Earth and strictly no sub-Neptunes (i.e., super-Earth-only systems) and of stars hosting at least one sub-Neptune and any number of super-Earths (i.e., any-sub-Neptune systems). Similarly to Section \ref{sec:discussion_perplanet}, we calculate the integrated signed CDF difference, and find that for all radius valley prescriptions considered, the integrated signed CDF difference is $>$ 0, demonstrating that the any-sub-Neptune hosts are typically more metal-rich than the super-Earth-only hosts. We observe a significant discrepancy between the super-Earth-only and any-sub-Neptune systems, and are able to reject the null hypothesis that their stellar metallicity distributions are drawn from the same underlying population most confidently under the PM25 and CM20 scenarios using AD tests. We are not able to confidently reject the null hypothesis under the V21 radius valley prescription. 

We attribute the discrepancy between the PM25/CM20 results and the V21 result to the differing parameter spaces for which the radius valley relations apply. The sample of M dwarfs used in V21 had a median stellar mass of 0.33 $\rm{M_\odot}$, which is lower than our sample's median mass of 0.49 $\rm{M_\odot}$. The radius valley is found at smaller radii for planets around low-mass stars \citep{Cloutier2020}. Therefore, the V21 radius valley definition may ``misclassify'' planets in our sample, which could lead to the statistically weaker distinction between the stellar metallicity distributions of the only-super-Earth and any-sub-Neptune host stars. Furthermore, we perform the same checks as before on the stellar mass, luminosity, and effective temperature distributions. The resulting p-values indicate that the super-Earth-only and any-sub-Neptune host stars are inconsistent with being drawn from the same population under the V21 radius valley definition. For these reasons, we do not consider the higher AD test statistic p-value distributions under the V21 radius valley definition to weaken the observed enhanced metallicity of any-sub-Neptune system host stars under the PM25 and CM20 radius valley definitions.

We observe an enhanced metallicity of sub-Neptune host stars compared to super-Earth-only host stars, and this result is in line with the picture of planet formation discussed in Section \ref{sec:discussion_perplanet}. If sub-Neptunes preferentially form in metal-rich disks, where there is sufficient solid and volatile content beyond the water ice line to form these water-rich planets, if any sub-Neptune is able to form in a given disk, this should not preclude super-Earths forming. However, if we observe a super-Earth-only system, we expect these systems to be more metal-poor than those containing sub-Neptunes. This is borne out in the CDFs computed on a per system basis in Fig. \ref{fig:cdf_persystem} (distinct from the per planet CDFs presented in Fig. \ref{fig:cdf}), albeit at a lower statistical significance than the per-planet result. For the CM20 radius valley definition, we can reject the null hypothesis for the stellar metallicity distributions with p-value $<0.05$ in $\sim$ half of the samples. This result thus moderately supports the proposed water-rich formation of sub-Neptunes around low-mass stars. We again note that this discussion is limited to the \textit{observed} planets in our sample, and that future, completeness-corrected studies are needed to probe the true metallicity distributions of super-Earth-only and sub-Neptune planetary systems.

\subsection{Planet Orbital Period}

We also consider the relationship between stellar metallicity and planet orbital period. \citet{Wanderley2025} found that planets with orbital periods $< 4.3$ d and radii $> 3 \rm{R_\oplus}$ preferentially orbit metal-rich M dwarfs. This trend is recovered in this work, but is not a clear result due to the small number (14) of planets with radii $> 3 \rm{R_\oplus}$ and periods $< 4.3$ d in our sample. Similar results for FGK stars \citep[e.g.][]{Mulders2016} suggest that metal-rich stars may have formed from protostellar disks with enhanced solid fractions, providing a larger reservoir of solids for close-in planet formation. This result, combined with an enhanced surface density of M dwarf disks within 0.5 AU \citep{Gaidos2017}, may explain the apparent abundance of close-in planets orbiting M dwarfs \citep{Ment2023}.

\subsection{Planet Radius for Giant Planets $\&$ Brown Dwarfs ($R_p > 9 R_\oplus$)}
The expanded sample discussed in this work contains six substellar objects with radii $ >9 \rm{R_\oplus}$, two of which orbit metal-poor stars: HATS-76 b (TIC 170849515, TOI 555, [Fe/H] = $-0.17 \pm$ 0.14 dex, \citealt{Gore2024}) and TOI-1278 B (TIC 163539739, [Fe/H] = $-0.18 \pm$ 0.09 dex, \citealt{Gore2024}).

While giant planets are typically rare around low-mass stars \citep{bryant+23}, they appear to exhibit the same stellar metallicity dependence as those orbiting FGK stars. Giant planets mostly orbit metal-rich low-mass stars \citep{Chachan2025, Gan2025}. This result is consistent with close-in giant planets forming via core accretion followed by a runaway gas accretion phase, which is enhanced in metal-rich protoplanetary disks \citep{Ida2004}. 

HATS-76 b was confirmed by \citet{jordan+22}, who reported a stellar metallicity of $0.29 \pm 0.13$ dex using NIR spectra from ARCoIRIS on the Blanco 4-m telescope \citep{Abbott2016}. This is $\sim 3\sigma$ discrepant from the metallicity reported in \citet{Gore2024} ([Fe/H] = $-0.17 \pm$ 0.14 dex). This may be due to the SNR of the spectra used to calculate metallicity in both cases; the relations presented in \citet{newton+14} were calculated using spectra with SNR $>$ 200, far exceeding the SNR of 40 spectrum used in \citet{jordan+22}. The spectrum of HATS-76 used in \citet{Gore2024} had an SNR of 80, also somewhat lower than the SNR $\sim$ 100 spectra used to develop the \citet{mann+13d} relations. The large discrepancy between the reported metallicities may thus be due to low spectral quality, or to systematic offsets between the methods used to measure metallicity.

Unlike giant planets, brown dwarfs do not form more readily around metal-rich stars, pointing to a distinct formation pathway such as gravitational instability \citep{boss97, chabrier14, bowler20}. \citet{Schlaufman2018} found that close-in (a $<$ 0.25 AU) objects with masses $> 10 \rm{M_J}$ orbit stars with a distinct stellar metallicity distribution from less massive giant planets. \citet{artigau+21} found that TOI-1278 B is in fact a transiting brown dwarf with a mass of 18.5 $\pm$ 0.5 $\rm{M_{J}}$ orbiting the MOV star TOI-1278, using the SPIRou instrument on the 3.6 m Canada-France-Hawaii Telescope \citep{Donati2020}. Given this context, our measured metallicity of -0.18 $\pm$ 0.09 dex is not surprising.

\section{Conclusions} \label{sec:conclusions}

In this work, we measured metallicities for \ntargets{} cool dwarf stars which host confirmed planets or planet candidates. By obtaining NIR spectra using IRTF/SpeX, we determined that the stars in our sample had metallicities ([Fe/H]) ranging from -0.6 to + 0.6 dex with a median metallicity of 0.0 dex. Combining our sample of newly characterized candidate and confirmed planet host stars with existing literature, we created a sample of \nplanetsottal{} planets and planet candidates orbiting \nStartotal{} cool, low-mass stars with homogeneously derived stellar parameters. We studied whether the M dwarfs hosting super-Earths and sub-Neptunes had distinct metallicity distributions. We found that cool dwarfs hosting sub-Neptunes were typically more metal-rich than those hosting super-Earths (median [Fe/H]$_{SN}$ = 0.05 dex, [Fe/H]$_{SE}$ = -0.14 dex, $\sigma_{\rm{[Fe/H]}}$ = 0.2 dex). This result is in line with sub-Neptunes around low-mass stars forming in water-rich environments beyond the snow line before migrating inwards to their observed locations. Protoplanetary disks around metal-rich M dwarfs have higher solid contents than metal-poor disks, providing both the inventory of ices and refractory elements needed to form sub-Neptunes in outer disk regions. 

\clearpage

\begin{acknowledgments}

The authors wish to recognize and acknowledge the very significant cultural role that the summit of Maunakea has within the indigenous Hawaiian community. We are most fortunate to have the opportunity to conduct observations from this sacred mountain which is now colonized land. We also thank E. Gillis,  N. Gromek, I. Malsky, R. Pudritz, and D. Weisserman for insightful discussions.

This research has made use of the Exoplanet Follow-up Observation Program \citep{exofop} website, which is operated by the California Institute of Technology, under contract with the National Aeronautics and Space Administration under the Exoplanet Exploration Program. This paper presented data collected by observing with the Infrared Telescope Facility (IRTF), which is operated by the University of Hawaii under contract 80HQTR24DA010 with the National Aeronautics and Space Administration. This paper made use of data collected by the \tess{} mission and are publicly available from the Mikulski Archive for Space Telescopes (MAST) operated by the Space Telescope Science Institute \citep{ticv82, tictoi}. Funding for the \tess{} mission is provided by NASA’s Science Mission Directorate. We acknowledge the use of public \tess{} data from pipelines at the \tess{} Science Office and at the \tess{} Science Processing Operations Center. Resources supporting this work were provided by the NASA High-End Computing (HEC) Program through the NASA Advanced Supercomputing (NAS) Division at Ames Research Center for the production of the SPOC data products. This research has made use of NASA’s Astrophysics Data System, and of the NASA Exoplanet Archive \citep{nea1}. 

SG is supported by an NSF Astronomy and Astrophysics Postdoctoral Fellowship under award AST-2303922.
\end{acknowledgments}

\vspace{5mm}
\facilities{IRTF/SpeX \citep{spex}, TESS \citep{Ricker2015}} 

\software{\texttt{numpy} \citep{harris2020array}, \texttt{astropy} \citep{astropy:2013, astropy:2018, astropy:2022}, \texttt{pandas} \citep{mckinney-proc-scipy-2010, reback2020pandas}, \texttt{astroquery} \citep{Ginsburg2019}, \texttt{lightkurve} \citep{Lightkurve_2018}, \texttt{tellrv} \citep{tellrv}, \texttt{matplotlib} \citep{Hunter:2007}, \texttt{exoplanet} \citep{exoplanet:agol20, exoplanet:arviz, exoplanet:joss, exoplanet:luger18, exoplanet:pymc3, exoplanet:theano, exoplanet:zenodo}, \texttt{Spextool} \citep{spextool, spextool2}}

\appendix
\section{Imaging Observations} \label{sec:appendix_imaging}
In this section we review the imaging observations of our targets, as executed by members of the \tess{} Follow-up Observing Program Working Group Sub-Group 3 (SG3). This sub-group conducts high-resolution imaging of TOIs, using adaptive optics, speckle imaging, and/or lucky imaging on a variety of instruments. The aim of SG3 is to to detect nearby objects that are not resolved in the \tess{} Input Catalog or by seeing-limited photometry. Of our sample, 34 targets have imaging observations. Several targets have multiple imaging observations; we make use of the observations taken in the reddest filter in each case. The spectral energy distributions of cool stars, such as those in our sample, are more similar at longer wavelengths than in the optical. As a result, their magnitudes are also more similar, and therefore more likely to be detectable in a given contrast curve. Each of the contrast curves that we use are available to download on ExoFOP-\tess{}\footnote{\url{https://exofop.ipac.caltech.edu/tess/}}, and we summarized the observations used in Table \ref{tab:ao_obs}.
\begin{table*}[ht]
\centering
\footnotesize
\caption{Summary of High-Resolution imaging observations used in Section \ref{sec:validation}. 
An asterisk (*) denotes the filter used in the vetting analysis when multiple were available. \label{tab:ao_obs}}
\begin{tabular}{llllllll}
\hline
TIC ID & TOI & Instrument & Telescope & Date & Filters & Contrast at 0.5\arcsec [mag] & Companions \\
\hline
\hline
415969908 & 233 & `Alopeke & Gemini-N & 12 September 2019 & 562, 832$^{*}$ nm & 4.95, 6.25& None \\
312862941 & 1638 & `Alopeke & Gemini-N & 03 December 2020 & 562, 832$^{*}$ nm & 4.68, 5.83& None \\
232635922 & 1745 & PHARO & Palomar & 24 August 2021 & \brg{}$^{*}$ & 5.877& $\Delta$mag = 3.75 $\pm$ 0.02,  \\
 & & & & & & & $\theta$ = 1.003\arcsec, PA = $10^\circ \pm 1^\circ$ \\
332477926 & 1754 & NESSI & WIYN & 25 April 2021 & 562, 832$^{*}$ nm & 4.372& None \\
138762614 & 1802 & NIRC2 & Keck-II & 28 May 2020 & \brg{} & 7.34& None \\
166648874 & 1806 & PHARO & Palomar & 05 December 2020 & \brg{} & 4.779& None \\
321669174 & 2081 & PHARO & Palomar & 19 March 2022 & \brg{} & 6.213& None \\
243313296 & 2278 & PHARO & Palomar & 24 August 2021 & H-cont, \brg{}$^{*}$ & 6.952, 6.968 & None \\
405763009 & 2433 & PHARO & Palomar & 19 September 2021 & \brg{} & 7.122 & None \\
59128183 & 2453 & PHARO & Palomar & 24 February 2021 & \brg{} & 6.968& None \\
202185707 & 4325 & PHARO & Palomar & 19 March 2022 & \brg{} & 6.169& None \\
337385330 & 5112 & PHARO & Palomar & 19 March 2022 & \brg{} & 6.434& None \\
389040826 & 5511 & Zorro & Gemini-S & 03 January 2024 & 562, 832$^{*}$ nm & 4.68, 5.74& None \\
443582629 & 5519 & NESSI & WIYN & 25 April 2021 & 562, 832$^{*}$ nm & 4.00, 4.84& None \\
422217860 & 5662 & PHARO & Palomar & 19 September 2021 & H-cont, K-cont$^{*}$ & 7.093, 6.716& None \\
310380289 & 5736 & NIRC2 & Keck-II & 10 June 2023 & K-cont & 7.345& None \\
262689575 & 5961 & Zorro & Gemini-S & 04 July 2023 & 562, 832$^{*}$ nm & 5.02, 6.55& None \\
354173595 & 5974 & NIRC2 & Keck-II & 05 August 2023 & J, K-cont$^{*}$ & 6.916, 7.860& None \\
120045750 & 5981 & PHARO & Palomar & 29 June 2023 & \brg{} & 7.129& None \\
284900292 & 5987 & PHARO & Palomar & 01 July 2023 & \brg{} & 6.62& None \\
196066560 & 6077 & PHARO & Palomar & 29 June 2023 & H-cont, \brg{}$^{*}$ & 6.280, 6.661& None \\
\hline
\end{tabular}%
\end{table*}

\section{Stellar \& Planet Parameters}
In this section we include Table \ref{tab:updated_params} detailing the stellar parameters measured using IRTF/SpeX spectra for the \ntois{} stars in our sample. We also include Tables \ref{tab:cpfits} and \ref{tab:pcfits}, which detail the planet parameters resulting from photometric fits for the confirmed planets and planet candidates in our sample, respectively.

\begin{longrotatetable}
\begin{deluxetable}{llllllllll}
\tabletypesize{\scriptsize}
\tablewidth{\textwidth}
\tablecaption{Updated stellar parameters using IRTF/SpeX spectra}
\tablehead{
\colhead{ID}&\colhead{TOI}&\colhead{Other Names}&\colhead{$R_\star$ ($R_\odot$)}&\colhead{$M_\star$ ($M_\odot$)}&\colhead{$L_\star$ ($L_\odot$)}&\colhead{log(g) (dex)}&\colhead{$T_{eff}$ (K)}&\colhead{[Fe/H] (dex)}&\colhead{[M/H] (dex)} \label{tab:updated_params}}
\startdata
 16005254 & 5344 &                                                & 0.57 $\pm$ 0.02 & 0.57 $\pm$ 0.02 &  0.0553 $\pm$ 0.002 & 4.68 $\pm$ 0.03 & 3704 $\pm$ 63 &  0.41 $\pm$ 0.09 &  0.24 $\pm$ 0.09 \\
 17307715 & 5562 &                                  K2-155, LP 415-17 & 0.59 $\pm$ 0.02 & 0.57 $\pm$ 0.02 & 0.0721 $\pm$ 0.0025 & 4.65 $\pm$ 0.03 & 3891 $\pm$ 64 & -0.57 $\pm$ 0.08 & -0.45 $\pm$ 0.08 \\
 18318288 & 6086 &                                                & 0.26 $\pm$ 0.01 & 0.23 $\pm$ 0.02 & 0.0064 $\pm$ 0.0002 & 4.97 $\pm$ 0.05 & 3215 $\pm$ 50 & -0.09 $\pm$ 0.08 & -0.09 $\pm$ 0.08 \\
 19028197 & 5094 &                                 GJ 3470, Kaewkosin\tablenotemark{a} & 0.49 $\pm$ 0.01 &  0.5 $\pm$ 0.02 & 0.0399 $\pm$ 0.0011 & 4.75 $\pm$ 0.03 & 3673 $\pm$ 58 &  0.25 $\pm$ 0.08 &  0.12 $\pm$ 0.08 \\
 26054627 & 5349 &                                                & 0.57 $\pm$ 0.02 & 0.58 $\pm$ 0.02 & 0.0607 $\pm$ 0.0021 & 4.69 $\pm$ 0.03 & 3796 $\pm$ 64 &  0.57 $\pm$ 0.11 &  0.27 $\pm$ 0.11 \\
 48353358 & 6004 &              KIC 10905746, KOI-1725, Kepler-1651 A & 0.49 $\pm$ 0.01 & 0.49 $\pm$ 0.02 & 0.0382 $\pm$ 0.0018 & 4.74 $\pm$ 0.03 & 3638 $\pm$ 66 & -0.05 $\pm$ 0.08 & -0.09 $\pm$ 0.08 \\
 59128183 & 2453 &                                                &  0.5 $\pm$ 0.01 &  0.5 $\pm$ 0.02 &  0.0398 $\pm$ 0.001 & 4.74 $\pm$ 0.03 & 3649 $\pm$ 56 &   0.1 $\pm$ 0.09 & -0.01 $\pm$ 0.09 \\
 67646988 & 1779 &                                                &  0.3 $\pm$ 0.01 & 0.28 $\pm$ 0.02 &  0.009 $\pm$ 0.0002 & 4.94 $\pm$ 0.04 & 3248 $\pm$ 50 &  0.57 $\pm$ 0.09 &  0.42 $\pm$ 0.09 \\
 77156829 &  696 &                                         LHS 1678 A & 0.35 $\pm$ 0.01 & 0.32 $\pm$ 0.02 & 0.0142 $\pm$ 0.0006 & 4.87 $\pm$ 0.04 & 3393 $\pm$ 62 & -0.45 $\pm$ 0.09 & -0.36 $\pm$ 0.08 \\
102734241 & 6002 &                                                & 0.24 $\pm$ 0.01 & 0.21 $\pm$ 0.02 & 0.0061 $\pm$ 0.0002 & 4.99 $\pm$ 0.05 & 3267 $\pm$ 56 & -0.03 $\pm$ 0.09 & -0.02 $\pm$ 0.09 \\
119584412 & 1801 &                               HIP 57099, LP 375-23 & 0.55 $\pm$ 0.02 & 0.54 $\pm$ 0.02 & 0.0538 $\pm$ 0.0018 &  4.7 $\pm$ 0.03 & 3760 $\pm$ 62 & -0.11 $\pm$ 0.08 & -0.14 $\pm$ 0.08 \\
120045750 & 5981 &                                        KIC 3932730 & 0.54 $\pm$ 0.01 & 0.54 $\pm$ 0.02 & 0.0452 $\pm$ 0.0015 & 4.71 $\pm$ 0.03 & 3631 $\pm$ 59 &  0.05 $\pm$ 0.08 &  0.05 $\pm$ 0.08 \\
126606859 & 4479 &                                                & 0.43 $\pm$ 0.01 & 0.42 $\pm$ 0.02 & 0.0249 $\pm$ 0.0009 &  4.8 $\pm$ 0.03 & 3495 $\pm$ 59 & -0.18 $\pm$ 0.09 & -0.15 $\pm$ 0.08 \\
138762614 & 1802 &                                                & 0.58 $\pm$ 0.02 & 0.57 $\pm$ 0.02 &  0.066 $\pm$ 0.0017 & 4.67 $\pm$ 0.03 & 3841 $\pm$ 59 &  0.01 $\pm$ 0.08 & -0.03 $\pm$ 0.08 \\
138819293 & 1796 &                GJ 436, HIP 57087, LHS 310, Noquisi\tablenotemark{b} & 0.42 $\pm$ 0.01 & 0.42 $\pm$ 0.02 &  0.024 $\pm$ 0.0007 & 4.81 $\pm$ 0.03 & 3488 $\pm$ 55 &  0.02 $\pm$ 0.08 & -0.03 $\pm$ 0.08 \\
166648874 & 1806 &                                                & 0.39 $\pm$ 0.01 & 0.39 $\pm$ 0.02 & 0.0183 $\pm$ 0.0005 & 4.84 $\pm$ 0.03 & 3390 $\pm$ 53 &   0.5 $\pm$ 0.09 &  0.35 $\pm$ 0.09 \\
196066560 & 6077 &                                                & 0.66 $\pm$ 0.02 & 0.63 $\pm$ 0.02 &  0.0951 $\pm$ 0.003 &  4.6 $\pm$ 0.03 & 3952 $\pm$ 63 & -0.13 $\pm$ 0.08 & -0.11 $\pm$ 0.08 \\
198385543 & 1846 &                                                & 0.42 $\pm$ 0.01 &  0.4 $\pm$ 0.02 & 0.0231 $\pm$ 0.0006 & 4.81 $\pm$ 0.03 & 3493 $\pm$ 53 & -0.22 $\pm$ 0.08 & -0.17 $\pm$ 0.08 \\
202185707 & 4325 &                                                & 0.45 $\pm$ 0.01 & 0.45 $\pm$ 0.02 & 0.0326 $\pm$ 0.0015 & 4.78 $\pm$ 0.03 & 3652 $\pm$ 67 &  0.11 $\pm$ 0.09 &  0.02 $\pm$ 0.09 \\
203289099 & 5111 &                             EPIC 212081533, K2-244 & 0.56 $\pm$ 0.02 & 0.55 $\pm$ 0.02 &  0.0595 $\pm$ 0.002 & 4.69 $\pm$ 0.03 & 3816 $\pm$ 63 &  0.06 $\pm$ 0.09 &   0.0 $\pm$ 0.09 \\
219041246 & 5713 &                                                & 0.29 $\pm$ 0.01 & 0.27 $\pm$ 0.02 & 0.0085 $\pm$ 0.0002 & 4.93 $\pm$ 0.04 & 3242 $\pm$ 49 &  0.02 $\pm$ 0.08 &  0.03 $\pm$ 0.08 \\
232635922 & 1745 &                                                & 0.43 $\pm$ 0.01 & 0.42 $\pm$ 0.02 & 0.0278 $\pm$ 0.0007 & 4.78 $\pm$ 0.03 & 3574 $\pm$ 55 & -0.56 $\pm$ 0.09 & -0.48 $\pm$ 0.09 \\
243313296 & 2278 &                                                & 0.33 $\pm$ 0.01 &  0.3 $\pm$ 0.02 & 0.0117 $\pm$ 0.0004 & 4.89 $\pm$ 0.04 & 3323 $\pm$ 56 & -0.23 $\pm$ 0.09 &  -0.2 $\pm$ 0.09 \\
262689575 & 5961 &                                         HIP 114941 & 0.62 $\pm$ 0.02 & 0.61 $\pm$ 0.02 & 0.0868 $\pm$ 0.0027 & 4.64 $\pm$ 0.03 & 3985 $\pm$ 63 &  0.09 $\pm$ 0.08 &  0.04 $\pm$ 0.08 \\
277833995 & 5524 &                             EPIC 248480671, K2-321 & 0.58 $\pm$ 0.02 & 0.58 $\pm$ 0.02 & 0.0635 $\pm$ 0.0019 & 4.68 $\pm$ 0.03 & 3815 $\pm$ 61 &  0.45 $\pm$ 0.09 &  0.31 $\pm$ 0.08 \\
284900292 & 5987 &                                                & 0.62 $\pm$ 0.02 & 0.61 $\pm$ 0.02 & 0.0823 $\pm$ 0.0033 & 4.64 $\pm$ 0.03 & 3927 $\pm$ 67 & -0.01 $\pm$ 0.09 & -0.06 $\pm$ 0.08 \\
307809773 & 4599 & BD+17 1320, GJ 239, HD 260655, HIP 31635, LHS 1858 & 0.47 $\pm$ 0.01 & 0.46 $\pm$ 0.02 & 0.0348 $\pm$ 0.0011 & 4.76 $\pm$ 0.03 & 3631 $\pm$ 59 & -0.32 $\pm$ 0.08 & -0.27 $\pm$ 0.08 \\
310380289 & 5736 &                                                & 0.58 $\pm$ 0.02 & 0.57 $\pm$ 0.02 & 0.0681 $\pm$ 0.0019 & 4.67 $\pm$ 0.03 & 3873 $\pm$ 61 & -0.08 $\pm$ 0.08 & -0.09 $\pm$ 0.08 \\
312862941 & 1638 &                                                & 0.69 $\pm$ 0.02 & 0.66 $\pm$ 0.02 & 0.1161 $\pm$ 0.0035 & 4.58 $\pm$ 0.03 & 4044 $\pm$ 64 &  0.12 $\pm$ 0.08 &  0.06 $\pm$ 0.08 \\
321669174 & 2081 &                                                & 0.52 $\pm$ 0.01 & 0.51 $\pm$ 0.02 &  0.045 $\pm$ 0.0012 & 4.72 $\pm$ 0.03 & 3688 $\pm$ 57 & -0.24 $\pm$ 0.08 & -0.22 $\pm$ 0.08 \\
332477926 & 1754 &                                                & 0.59 $\pm$ 0.02 & 0.58 $\pm$ 0.02 & 0.0723 $\pm$ 0.0031 & 4.66 $\pm$ 0.03 & 3887 $\pm$ 69 & -0.06 $\pm$ 0.08 &  -0.1 $\pm$ 0.08 \\
337385330 & 5112 &                                                & 0.61 $\pm$ 0.02 & 0.59 $\pm$ 0.02 & 0.0848 $\pm$ 0.0022 & 4.65 $\pm$ 0.03 & 4000 $\pm$ 62 &  -0.1 $\pm$ 0.09 & -0.11 $\pm$ 0.09 \\
348755728 & 1883 &                                                & 0.49 $\pm$ 0.01 & 0.49 $\pm$ 0.02 &   0.034 $\pm$ 0.001 & 4.75 $\pm$ 0.03 & 3537 $\pm$ 58 &   0.08 $\pm$ 0.1 &  0.03 $\pm$ 0.09 \\
354173595 & 5974 &                                                & 0.37 $\pm$ 0.01 & 0.36 $\pm$ 0.02 & 0.0166 $\pm$ 0.0005 & 4.85 $\pm$ 0.03 & 3382 $\pm$ 53 & -0.18 $\pm$ 0.08 & -0.19 $\pm$ 0.08 \\
368287008 & 2015 &                                                & 0.32 $\pm$ 0.01 &  0.3 $\pm$ 0.02 & 0.0104 $\pm$ 0.0003 & 4.91 $\pm$ 0.04 & 3250 $\pm$ 51 &  0.24 $\pm$ 0.08 &  0.19 $\pm$ 0.08 \\
387974148 & 5551 &                    Gar\tablenotemark{c}, GJ 486, HIP 62452, LHS 341 & 0.55 $\pm$ 0.02 & 0.55 $\pm$ 0.02 & 0.0566 $\pm$ 0.0017 &  4.7 $\pm$ 0.03 & 3798 $\pm$ 60 &  0.06 $\pm$ 0.09 &  0.01 $\pm$ 0.08 \\
389040826 & 5511 &                                                & 0.51 $\pm$ 0.01 & 0.51 $\pm$ 0.02 &  0.044 $\pm$ 0.0013 & 4.73 $\pm$ 0.03 & 3695 $\pm$ 59 &  0.05 $\pm$ 0.09 &  0.01 $\pm$ 0.09 \\
390651552 & 1827 &                                                & 0.33 $\pm$ 0.01 & 0.31 $\pm$ 0.02 & 0.0119 $\pm$ 0.0003 & 4.89 $\pm$ 0.04 & 3306 $\pm$ 52 & -0.03 $\pm$ 0.08 & -0.04 $\pm$ 0.08 \\
405763009 & 2433 &                                                & 0.31 $\pm$ 0.01 & 0.29 $\pm$ 0.02 & 0.0099 $\pm$ 0.0002 & 4.91 $\pm$ 0.04 & 3268 $\pm$ 50 & -0.06 $\pm$ 0.08 & -0.06 $\pm$ 0.08 \\
407591297 & 5388 &                                                & 0.32 $\pm$ 0.01 & 0.29 $\pm$ 0.02 & 0.0109 $\pm$ 0.0002 &  4.9 $\pm$ 0.04 & 3309 $\pm$ 50 &  -0.3 $\pm$ 0.08 & -0.26 $\pm$ 0.08 \\
417931300 & 2068 &                                                & 0.54 $\pm$ 0.02 & 0.54 $\pm$ 0.02 & 0.0504 $\pm$ 0.0012 &  4.7 $\pm$ 0.03 & 3717 $\pm$ 57 &  -0.0 $\pm$ 0.08 & -0.02 $\pm$ 0.08 \\
422217860 & 5662 &                                                &  0.6 $\pm$ 0.02 & 0.61 $\pm$ 0.02 &  0.076 $\pm$ 0.0023 & 4.66 $\pm$ 0.03 & 3901 $\pm$ 63 &  0.56 $\pm$ 0.08 &  0.38 $\pm$ 0.08 \\
434116397 & 5955 &                                         HIP 116907 & 0.59 $\pm$ 0.02 & 0.58 $\pm$ 0.02 & 0.0706 $\pm$ 0.0021 & 4.66 $\pm$ 0.03 & 3880 $\pm$ 61 &  0.03 $\pm$ 0.08 & -0.03 $\pm$ 0.08 \\
434226736 & 5095 &                              EPIC 210490365, K2-25 & 0.29 $\pm$ 0.01 & 0.27 $\pm$ 0.02 & 0.0079 $\pm$ 0.0002 & 4.94 $\pm$ 0.04 & 3192 $\pm$ 51 &  0.07 $\pm$ 0.08 & -0.03 $\pm$ 0.08 \\
443582629 & 5519 &                                                & 0.37 $\pm$ 0.01 & 0.35 $\pm$ 0.02 & 0.0162 $\pm$ 0.0005 & 4.85 $\pm$ 0.04 & 3394 $\pm$ 56 & -0.15 $\pm$ 0.08 & -0.16 $\pm$ 0.08 \\
441420236 & 2221 & HD 197481, AU Mic, GJ 803, HIP 102409 & 0.68 $\pm$ 0.02 & 0.66 $\pm$ 0.02 & 0.0978 $\pm$ 0.0024 & 4.59 $\pm$ 0.03 &  3906 $\pm$ 60 &  0.11 $\pm$ 0.08 &  0.07 $\pm$ 0.08 \\
388804061 & 5555 &                                 K2-18 & 0.44 $\pm$ 0.01 & 0.44 $\pm$ 0.02 & 0.0264 $\pm$ 0.0008 & 4.79 $\pm$ 0.03 &  3501 $\pm$ 55 &   0.1 $\pm$ 0.08 &  0.05 $\pm$ 0.08 \\
173103335 & 5146 &                                  K2-3 & 0.56 $\pm$ 0.02 & 0.55 $\pm$ 0.02 &  0.056 $\pm$ 0.0016 & 4.69 $\pm$ 0.03 &  3767 $\pm$ 59 & -0.26 $\pm$ 0.08 & -0.24 $\pm$ 0.08 \\
 38337202 &     - &                                           K2-72 & 0.33 $\pm$ 0.01 & 0.31 $\pm$ 0.02 & 0.0135 $\pm$ 0.0004 & 4.88 $\pm$ 0.04 &  3400 $\pm$ 55 &  -0.39 $\pm$ 0.1 &   -0.3 $\pm$ 0.1 \\
 82050863 &     - &                                            K2-9 & 0.33 $\pm$ 0.01 &  0.3 $\pm$ 0.02 & 0.0109 $\pm$ 0.0004 & 4.89 $\pm$ 0.04 &  3266 $\pm$ 57 & -0.22 $\pm$ 0.09 & -0.18 $\pm$ 0.09 \\
 27187450 &     - &                            KOI 2626,Kepler-1652 & 0.41 $\pm$ 0.05 & 0.39 $\pm$ 0.05 & 0.0218 $\pm$ 0.0073 & 4.82 $\pm$ 0.11 & 3481 $\pm$ 351 &   -0.2 $\pm$ 0.1 &  -0.16 $\pm$ 0.1 \\
239275865 &     - &                           KOI 463, Kepler 560 B & 0.41 $\pm$ 0.01 &  0.4 $\pm$ 0.02 & 0.0222 $\pm$ 0.0008 & 4.81 $\pm$ 0.03 &  3474 $\pm$ 59 & -0.31 $\pm$ 0.09 & -0.25 $\pm$ 0.09 \\
268159861 &     - &                             KOI 571, Kepler 186 & 0.55 $\pm$ 0.02 & 0.54 $\pm$ 0.02 & 0.0556 $\pm$ 0.0018 & 4.69 $\pm$ 0.03 &  3769 $\pm$ 61 & -0.26 $\pm$ 0.08 & -0.21 $\pm$ 0.08 \\
273373582 &     - &                           KOI 2418, Kepler 1229 & 0.56 $\pm$ 0.02 & 0.56 $\pm$ 0.02 & 0.0601 $\pm$ 0.0024 & 4.68 $\pm$ 0.03 &  3807 $\pm$ 67 &   0.02 $\pm$ 0.1 &  -0.04 $\pm$ 0.1 \\
122450207 &     - &                           KOI 3010, Kepler 1410 & 0.52 $\pm$ 0.03 & 0.52 $\pm$ 0.03 & 0.0505 $\pm$ 0.0085 & 4.71 $\pm$ 0.06 & 3778 $\pm$ 199 & -0.12 $\pm$ 0.12 & -0.07 $\pm$ 0.12 \\
158989438 &     - &                           KOI 3497, Kepler 1512 & 0.34 $\pm$ 0.05 & 0.31 $\pm$ 0.06 & 0.0191 $\pm$ 0.0071 & 4.88 $\pm$ 0.16 & 3708 $\pm$ 443 & -0.13 $\pm$ 0.09 & -0.14 $\pm$ 0.09 \\
 28090925 &     - &                           KOI 4036, Kepler 1544 & 0.76 $\pm$ 0.02 &  0.7 $\pm$ 0.02 &  0.235 $\pm$ 0.0088 & 4.52 $\pm$ 0.03 &  4605 $\pm$ 77 & -0.36 $\pm$ 0.09 &  -0.3 $\pm$ 0.09 \\
120499135 &     - &                            KOI 4742, Kepler 442 & 0.69 $\pm$ 0.02 & 0.64 $\pm$ 0.02 &  0.1301 $\pm$ 0.004 & 4.57 $\pm$ 0.03 &  4179 $\pm$ 68 & -0.58 $\pm$ 0.09 & -0.43 $\pm$ 0.09 \\
415969908 &  233 &                              LP 821-31 & 0.38 $\pm$ 0.01 & 0.37 $\pm$ 0.02 & 0.0182 $\pm$ 0.0005 & 4.84 $\pm$ 0.03 &  3424 $\pm$ 53 & -0.21 $\pm$ 0.08 & -0.18 $\pm$ 0.08 
\enddata
\vspace{0.5em}
\tablenotemark{a}{Name assigned by the IAU NameExoWorlds Program. ``Kaewkosin" refers to the crystals of the Hindu deity Indra in the Thai language.} \\
\tablenotemark{b}{Name assigned by the IAU NameExoWorlds Program. ``Noquisi" means ``star" in the Cherokee language.} \\ 
\tablenotemark{c}{Name assigned by the IAU NameExoWorlds Program. ``Gar" means ``flame" in the Basque language.}
\end{deluxetable}
\end{longrotatetable}
\clearpage

\begin{longrotatetable}
\begin{longdeluxetable}{llllllllllllcl}
\tabletypesize{\scriptsize}
\tablewidth{\textwidth}
\tablecaption{Updated planet parameters for confirmed planets}
\tablehead{\colhead{TIC} & \colhead{Planet} & \colhead{$\rm{P_{orb}}$} & \colhead{$\rm{\sigma_{P}}$} & \colhead{$\rm{T_{c}}$} & \colhead{$\rm{\sigma_{T_{c}}}$}  & \colhead{b}  & \colhead{$\rm{\sigma_b}$} & \colhead{$\rm{{R_p}/{R_*}}$} & \colhead{$\rm{\sigma_{{R_p}/{R_*}}}$} & \colhead{$\rm{{R_p}}$}  & \colhead{$\rm{\sigma_{R_p}}$}  & \colhead{Stellar}  &  \colhead{Planet}
\\
 &  \colhead{Name}  &  \colhead{(d)}  &  \colhead{(d)}  & \colhead{(BJD)} &  \colhead{(BJD)}  &  &  &  &  & \colhead{($\rm{R_\oplus}$)} &  \colhead{($\rm{R_\oplus}$)}  & \colhead{Reference}   & \colhead{Reference} \label{tab:cpfits}} 
\startdata
203289099 &    * K2-344 b &          3.3558500 &             0.0000920 &                2458095.74773 &                     0.00126 & 0.410 &           0.305 &             0.0295 &                      0.0014 &         1.79 &                0.13 &         This Work &               \cite{deLeon2021}  \\
138819293 &      GJ 436 b &          2.6438831 &             0.0000006 &                2454510.80162 &                     0.00007 & 0.736 &           0.042 &             0.0822 &                      0.0011 &         3.81 &                0.12 &         This Work &                            \cite{Maciejewski2014b} \\
 434226736 &       K2-25 b &          3.4845641 &             0.0000006 &                2458515.64206 &                     0.00010 & 0.628 &           0.035 &             0.1075 &                      0.0018 &         3.40 &                0.11 &         This Work &                             \cite{Stefansson2020b} \\
 17307715 &      K2-155 b &          6.3420000 &             0.0020000 &                2457818.71530 &                     0.00210 & 0.340 &           0.235 &             0.0261 &                      0.0009 &         1.68 &                0.07 &         This Work &                             \cite{DiezAlonso2018a} \\
 17307715 &      K2-155 c &         13.8500000 &             0.0060000 &                2457814.56430 &                     0.00250 & 0.450 &           0.340 &             0.0321 &                      0.0013 &         2.07 &                0.10 &         This Work &                             \cite{DiezAlonso2018a} \\
 17307715 &      K2-155 d &         40.7180000 &             0.0050000 &                2457782.83240 &                     0.00480 & 0.440 &           0.315 &             0.0273 &                      0.0018 &         1.76 &                0.13 &         This Work &                             \cite{DiezAlonso2018a} \\
277833995 &    * K2-321 b &          2.2979749 &             0.0000018 &                2458141.26759 &                     0.00066 & 0.470 &           0.310 &             0.0315 &                      0.0017 &         1.98 &                0.14 &         This Work &       \cite{CastroGonzalez2020}  \\
390651552 &      GJ 486 b &          1.4671213 &             0.0000003 &                2459939.07160 &                     0.00001 & 0.120 &           0.081 &             0.0372 &                      0.0001 &         1.35 &                0.04 &         This Work &                               \cite{Trifonov2021a} \\
 77156829 &    LHS 1678 b &          0.8602325 &             0.0000012 &                2458998.15553 &                     0.00071 & 0.210 &           0.150 &             0.0191 &                      0.0008 &         0.72 &                0.04 &         This Work &                             \cite{Silverstein2022} \\
 77156829 &    LHS 1678 c &          3.6942840 &             0.0000046 &                2458998.45607 &                     0.00061 & 0.440 &           0.115 &             0.0262 &                      0.0011 &         0.99 &                0.05 &         This Work &                             \cite{Silverstein2022} \\
307809773 &   HD 260655 b &          2.7695300 &             0.0000300 &                2459497.91020 &                     0.00030 & 0.665 &           0.023 &             0.0259 &                      0.0005 &         1.33 &                0.04 &         This Work &                                  \cite{Luque2022a} \\
307809773 &   HD 260655 c &          5.7058800 &             0.0000700 &                2459490.36460 &                     0.00040 & 0.890 &           0.007 &             0.0320 &                      0.0010 &         1.65 &                0.07 &         This Work &                                  \cite{Luque2022a} \\
126606859 &  * TOI-4479 b &          1.1589000 &             0.0000150 &                2459420.75780 &                     0.00120 & 0.430 &           0.270 &             0.0620 &                      0.0075 &         2.91 &                0.56 &         This Work &      \cite{Esparza-Borges2022b}  \\
321669174 &  * TOI-2081 b &         10.5053400 &             0.0000750 &                2458685.89960 &                     0.00285 & 0.350 &           0.255 &             0.0396 &                      0.0032 &         2.25 &                0.26 &         This Work &      \cite{Esparza-Borges2022b}  \\
119584412 &    TOI-1801 b &         10.6438700 &             0.0000550 &                2458903.54351 &                     0.00328 & 0.265 &           0.180 &             0.0340 &                      0.0010 &         2.03 &                0.08 &         This Work &                             \cite{Mallorquin2023b} \\
166648874 &   TOI-1806.01 &         15.1454693 &                   - &                          - &                         - &   - &             - &             0.0778 &                         - &         3.33 &                 - &         This Work &                                         \cite{Hord2024} \\
198385543 &    TOI-1846 b &          3.9306737 &             0.0000043 &                2459565.95523 &                     0.00031 & 0.464 &           0.051 &             0.0414 &                      0.0007 &         1.87 &                0.06 &         This Work &                                         \cite{Soubkiou2025} \\
348755728 &    TOI-1883 b &          4.5063800 &             0.0000065 &                2459256.84810 &                     0.00020 & 0.210 &           0.145 &             0.1076 &                      0.0029 &         5.76 &                0.28 &         This Work &                                         \cite{Pelaez-Torres2024} \\
368287008 &    TOI-2015 b &          3.3482370 &             0.0000325 &                2459424.78570 &                     0.00015 & 0.743 &           0.003 &             0.0928 &                      0.0003 &         3.25 &                0.09 &         This Work &                                         \cite{Jones2024} \\
417931300 &  * TOI-2068 b &          7.7689150 &             0.0000310 &                2458683.42580 &                     0.00265 & 0.460 &           0.280 &             0.0310 &                      0.0020 &         1.83 &                0.13 &         This Work &               \cite{Mistry2024}  \\
 59128183 &    TOI-2453 b &          4.4414346 &             0.0000129 &                2459494.61240 &                     0.00129 & 0.681 &           0.119 &             0.0590 &                      0.0035 &         3.21 &                0.24 &         This Work &                                         \cite{Lafarga2026} \\
 16005254 &    TOI-5344 b &          3.7926220 &             0.0000062 &                2459848.99030 &                     0.00019 & 0.735 &           0.010 &             0.1653 &                      0.0014 &        10.29 &                0.31 &         This Work &                                 \cite{Hartman2023} \\
 26054627 &    TOI-5349 b &          3.3179210 &             0.0000020 &                2459521.81840 &                     0.00050 & 0.510 &           0.040 &             0.1610 &                      0.0020 &        10.00 &                0.31 &         This Work &                                         \cite{Sandoval2026} \\
407591297 &   TOI-5388.01 &          2.5946748 &             0.0000036 &                2459348.07817 &                     0.00053 & 0.430 &           0.285 &             0.0305 &                      0.0014 &         1.06 &                0.07 &         This Work &                                        \cite{Poultourtzidis2026} \\
 77156829 &  * LHS 1678 d &          4.9652229 &             0.0000085 &                2459000.45806 &                     0.00088 & 0.760 &           0.050 &             0.0273 &                      0.0017 &         1.03 &                0.07 &         This Work &          \cite{Silverstein2024}  \\
219041246 &    TOI-5713 b &         10.4419890 &             0.0000145 &                2458745.67760 &                     0.00160 & 0.440 &           0.240 &             0.0541 &                      0.0046 &         1.73 &                0.17 &         This Work &                                         \cite{Ghachoui2024} \\
310380289 &    TOI-5736 b &          0.6489900 &             0.0000100 &                2458738.89683 &                     0.00149 & 0.435 &           0.194 &             0.0247 &                      0.0007 &         1.56 &                0.06 &         This Work &                                         \cite{GomezBarrientos2026} \\
102734241 &    TOI-6002 b &         10.9048210 &             0.0000195 &                2458692.76390 &                     0.00260 & 0.260 &           0.175 &             0.0622 &                      0.0075 &         1.65 &                0.22 &         This Work &                                         \cite{GomezBarrientos2026} \\
 18318288 &    TOI-6086 b &          1.3888725 &             0.0000827 &                2460131.96950 &                     0.00057 & 0.608 &           0.051 &             0.0417 &                      0.0022 &         1.18 &                0.07 &         This Work &                                         \cite{Barkaoui2024} \\
 48353358 & Kepler-1651 b &          9.8786392 &             0.0000104 &                2454961.53395 &                     0.00130 & 0.450 &           0.215 &             0.0336 &                      0.0009 &         1.80 &                0.07 &         This Work &                                   \cite{Mann2017b} \\
368287008 &    TOI-2015 c &          5.5827960 &             0.0000415 &                          - &                         - & 2.020 &           0.105 &                - &                         - &          - &                 - &         This Work &                                         \cite{Barkaoui2025} \\
441420236 &      AU Mic b &          8.4634460 &             0.0000050 &                2458330.35088 &                     0.00052 & 0.502 &           0.046 &             0.0499 &                      0.0004 &         3.80 &                0.34 &               This Work &                                         \cite{Mallorquin2024} \\
441420236 &      AU Mic c &         18.8590230 &             0.0000230 &                2458342.22333 &                     0.00110 & 0.259 &           0.191 &             0.0291 &                      0.0004 &         2.21 &                0.20 &               This Work &                                         \cite{Mallorquin2024} \\
388804061 &       K2-18 b &         32.9396230 &             0.0000975 &                2457264.39144 &                     0.00065 &   - &             - &             0.0529 &                      0.0069 &         2.55 &                0.34 &               This Work &                                         \cite{Sarkis2018} \\
 38337202 &       K2-72 b &          5.5772120 &             0.0004180 &                2457010.37600 &                     0.00200 &   - &             - &             0.0299 &                      0.0014 &         1.09 &                0.15 &               This Work &                                         \cite{Dressing2017b} \\
 38337202 &       K2-72 c &         15.1890340 &             0.0031385 &                2456989.46500 &                     0.00500 &   - &             - &             0.0323 &                      0.0021 &         1.17 &                0.17 &               This Work &                                         \cite{Dressing2017b} \\
 38337202 &       K2-72 d &          7.7601780 &             0.0014960 &                2456984.78800 &                     0.00750 &   - &             - &             0.0279 &                      0.0021 &         1.02 &                0.16 &               This Work &                                         \cite{Dressing2017b} \\
 38337202 &       K2-72 e &         24.1588680 &             0.0037880 &                2456987.05400 &                     0.00500 &   - &             - &             0.0358 &                      0.0019 &         1.31 &                0.19 &               This Work &                                         \cite{Dressing2017b} \\
 27187450 & Kepler-1652 b &         38.0972200 &             0.0002100 &                2454973.00450 &                     0.00470 & 0.011 &           0.122 &             0.0387 &                      0.0017 &         1.70 &                0.31 &               This Work &                                         \cite{Torres2017} \\
239275865 &  Kepler-560 b &         18.4776445 &             0.0000152 &                2455018.26856 &                     0.00061 &   - &             - &             0.0479 &                      0.0007 &         2.15 &                0.16 &               This Work &                                         \cite{Morton2016} \\
268159861 &  Kepler-186 b &          3.8867907 &             0.0000062 &                2454966.33040 &                     0.00130 & 0.300 &           0.200 &             0.0208 &                      0.0005 &         1.25 &                0.20 &               This Work &                                         \cite{Quintana2014} \\
268159861 &  Kepler-186 e &         22.4077040 &             0.0000730 &                2454986.80060 &                     0.00240 & 0.310 &           0.200 &             0.0246 &                      0.0006 &         1.49 &                0.24 &               This Work &                                         \cite{Quintana2014} \\
268159861 &  Kepler-186 f &        129.9441000 &             0.0012500 &                2455789.49400 &                     0.00385 & 0.130 &           0.205 &             0.0205 &                      0.0012 &         1.24 &                0.11 &               This Work &                                         \cite{Torres2015} \\
268159861 &  Kepler-186 c &          7.2673020 &             0.0000115 &                2455007.31420 &                     0.00125 & 0.280 &           0.195 &             0.0242 &                      0.0005 &         1.46 &                0.23 &               This Work &                                         \cite{Quintana2014} \\
268159861 &  Kepler-186 d &         13.3429960 &             0.0000245 &                2455009.90450 &                     0.00145 & 0.360 &           0.215 &             0.0272 &                      0.0007 &         1.64 &                0.26 &               This Work &                                         \cite{Quintana2014} \\
273373582 & Kepler-1229 b &         86.8289890 &             0.0010690 &                2455022.26884 &                     0.00777 &   - &             - &             0.0247 &                      0.0016 &         1.55 &                0.15 &               This Work &                                         \cite{Morton2016} \\
122450207 & Kepler-1410 b &         60.8661680 &             0.0005161 &                2455012.28498 &                     0.00688 &   - &             - &             0.0274 &                      0.0012 &         1.56 &                0.17 &               This Work &                                         \cite{Morton2016} \\
158989438 & Kepler-1512 b &         20.3597260 &             0.0000587 &                2454967.29096 &                     0.00240 &   - &             - &             0.0162 &                      0.0006 &         0.59 &                0.10 &               This Work &                                         \cite{Morton2016} \\
 28090925 & Kepler-1544 b &        168.8111740 &             0.0012710 &                2455044.44085 &                     0.00531 &   - &             - &             0.0220 &                      0.0008 &         1.83 &                0.11 &               This Work &                                         \cite{Morton2016} \\
120499135 &  Kepler-442 b &        112.3053000 &             0.0026000 &                2455849.55780 &                     0.00615 & 0.220 &           0.255 &             0.0211 &                      0.0018 &         1.54 &                0.15 &               This Work &                                         \cite{Torres2015} \\
173103335 &        K2-3 b &         10.0546535 &             0.0000090 &                2457165.32947 &                     0.00025 & 0.310 &           0.155 &             0.0349 &                      0.0017 &         2.17 &                0.16 & \tablenotemark{a} &                            \cite{Diamond-Lowe2022} \\
150096001 &    * K2-133 b &          3.0713300 &             0.0001100 &                2457821.31500 &                     0.00155 & 0.630 &           0.053 &             0.0270 &                      0.0020 &         1.39 &                0.13 & \tablenotemark{a} &               \cite{Wells2019}   \\
296739893 &   * TOI-620 b &          5.0988179 &             0.0000046 &                2458992.19724 &                     0.00073 & 0.887 &           0.015 &             0.0627 &                      0.0032 &         3.69 &                0.21 & \tablenotemark{b} &               \cite{Reefe2022a}  \\
144700903 &     TOI-532 b &          2.3266508 &             0.0000030 &                2458470.57678 &                     0.00088 &   - &             - &             0.0872 &                      0.0034 &         5.79 &                0.28 & \tablenotemark{b} &                                \cite{Kanodia2021b} \\
374180079 &    * K2-266 c &          7.8140000 &             0.0017500 &                2457946.65100 &                     0.00640 & 0.600 &           0.210 &             0.0092 &                      0.0013 &         0.64 &                0.09 & \tablenotemark{a} &           \cite{Rodriguez2018c}  \\
374339566 &    * K2-239 d &         10.1150000 &             0.0010000 &                2457908.38100 &                         - & 0.390 &           0.250 &             0.0280 &                      0.0027 &         0.83 &                0.17 & \tablenotemark{a} &          \cite{DiezAlonso2018b}  \\
 82050863 &      * K2-9 b &         18.4498000 &             0.0015000 &                2456822.67120 &                     0.00290 & 0.926 &           0.250 &             0.0665 &                      0.0283 &         2.69 &                1.18 & \tablenotemark{a} &           \cite{Schlieder2016b}  \\
218795833 &     TOI-519 b &          1.2652328 &             0.0000005 &                2458491.87712 &                         - &   - &             - &             0.3024 &                      0.0123 &        11.58 &                0.58 & \tablenotemark{b} &                                \cite{Kagetani2023} \\
150096001 &    * K2-133 d &         11.0245400 &             0.0003550 &                2457826.17090 &                     0.00120 & 0.370 &           0.060 &             0.0404 &                      0.0029 &         2.07 &                0.20 & \tablenotemark{a} &               \cite{Wells2019}   \\
283722336 &   HD 219134 d &         46.8590000 &             0.0280000 &                2455208.44000 &                         - &   - &             - &             0.0190 &                      0.0003 &         1.61 &                0.03 & \tablenotemark{d} &                                 \cite{Gillon2017b} \\
283722336 &   HD 219134 c &          6.7645800 &             0.0003300 &                2457474.04591 &                     0.00088 & 0.813 &           0.025 &             0.0178 &                      0.0006 &         1.51 &                0.05 & \tablenotemark{d} &                                 \cite{Gillon2017b} \\
374180079 &    * K2-266 b &          0.6585240 &             0.0000170 &                2457949.67470 &                     0.00320 & 1.011 &           0.026 &             0.0430 &                      0.0235 &         2.99 &                1.64 & \tablenotemark{a} &           \cite{Rodriguez2018c}  \\
 70899085 &   LP 714-47 b &          4.0520370 &             0.0000040 &                2458774.70336 &                         - &   - &             - &             0.0738 &                      0.0051 &         4.76 &                0.36 & \tablenotemark{b} &                                \cite{Dreizler2020} \\
408636441 &    TOI-1759 b &         18.8501900 &             0.0001300 &                2458745.46540 &                     0.00110 & 0.210 &           0.095 &             0.0482 &                      0.0020 &         3.27 &                0.16 & \tablenotemark{b} &                                \cite{Espinoza2022} \\
374339566 &    * K2-239 c &          7.7750000 &             0.0010000 &                2457916.86000 &                         - & 0.730 &           0.315 &             0.0255 &                      0.0026 &         0.75 &                0.16 & \tablenotemark{a} &          \cite{DiezAlonso2018b}  \\
 29960110 &    TOI-1201 b &          2.4919863 &             0.0000031 &                2459169.23222 &                     0.00053 & 0.404 &           0.076 &             0.0436 &                      0.0021 &         2.29 &                0.13 & \tablenotemark{b} &                             \cite{Kossakowski2021} \\
243185500 &    TOI-1468 c &         15.5324770 &             0.0000255 &                2458766.92670 &                     0.00120 & 0.664 &           0.145 &             0.0525 &                      0.0027 &         2.11 &                0.12 & \tablenotemark{b} &      \cite{MeierValdes2025, MeierValdes2025errata} \\
  5882269 &    * K2-284 b &          4.7950690 &             0.0000860 &                2457859.11316 &                     0.00043 & 0.280 &           0.190 &             0.0420 &                      0.0026 &         2.80 &                0.22 & \tablenotemark{a} &              \cite{David2018b}   \\
150096001 &    * K2-133 e &         26.5841000 &             0.0017500 &                2457837.86570 &                     0.00225 & 0.928 &           0.047 &             0.0349 &                      0.0033 &         1.79 &                0.20 & \tablenotemark{a} &               \cite{Wells2019}   \\
170849515 &     HATS-76 b &          1.9416423 &             0.0000014 &                2458424.55556 &                     0.00053 & 0.281 &           0.092 &             0.1772 &                      0.0056 &        12.48 &                0.55 & \tablenotemark{b} &                                  \cite{Jordan2022} \\
 83181153 &     * K2-26 b &         14.5665000 &             0.0018000 &                2456775.16590 &                     0.00245 & 0.570 &           0.315 &             0.0471 &                      0.0109 &         2.68 &                0.64 & \tablenotemark{a} &           \cite{Schlieder2016b}  \\
173103335 &        K2-3 d &         44.5560300 &             0.0001250 &                2457271.78796 &                     0.00076 & 0.240 &           0.160 &             0.0245 &                      0.0012 &         1.52 &                0.11 & \tablenotemark{a} &                            \cite{Diamond-Lowe2022} \\
437054764 &    K2-419 A b &         20.3584725 &             0.0000059 &                2459553.71380 &                     0.00050 & 0.389 &           0.089 &             0.1788 &                      0.0087 &        10.67 &                0.77 & \tablenotemark{c} &                                \cite{Kanodia2024b} \\
283722336 &   HD 219134 f &         22.7170000 &             0.0150000 &                2455233.17000 &                         - &   - &             - &             0.0154 &                      0.0003 &         1.31 &                0.02 & \tablenotemark{d} &                                 \cite{Gillon2017b} \\
 28900646 &    TOI-1685 b &          0.6691392 &             0.0000004 &                2459910.93828 &                     0.00040 & 0.266 &           0.069 &             0.0295 &                      0.0013 &         1.48 &                0.08 & \tablenotemark{b} &                                    \cite{Burt2024} \\
283722336 &   HD 219134 b &          3.0929260 &             0.0000100 &                2457126.69913 &                     0.00087 & 0.924 &           0.006 &             0.0189 &                      0.0007 &         1.60 &                0.06 & \tablenotemark{d} &                                 \cite{Gillon2017b} \\
203214081 &      G 9-40 b &          5.7459982 &             0.0000020 &                2459503.32682 &                     0.00042 & 0.513 &           0.042 &             0.0576 &                      0.0027 &         1.99 &                0.11 & \tablenotemark{b} &                                  \cite{Luque2022c} \\
389900760 &  * TOI-2120 b &          5.7998164 &             0.0000035 &                2458795.82368 &                     0.00041 & 0.550 &           0.085 &             0.0794 &                      0.0031 &         2.08 &                0.10 & \tablenotemark{b} &  \cite{Hori2024}  \\
423358488 &    * K2-148 b &          4.3839500 &             0.0008000 &                2457390.05956 &                     0.00961 &   - &             - &             0.0193 &                      0.0034 &         1.29 &                0.24 & \tablenotemark{a} &              \cite{Hirano2018b}  \\
 18310799 &      K2-136 c &         17.3071300 &             0.0002700 &                2457812.71770 &                     0.00085 & 0.310 &           0.125 &             0.0406 &                      0.0024 &         2.72 &                0.20 & \tablenotemark{a} &                                  \cite{Mayo2023} \\
423358488 &    * K2-148 d &          9.7579000 &             0.0010000 &                2457386.34305 &                     0.00545 &   - &             - &             0.0238 &                      0.0042 &         1.60 &                0.29 & \tablenotemark{a} &              \cite{Hirano2018b}  \\
374180079 &    * K2-266 e &         19.4820000 &             0.0012000 &                2457938.54100 &                     0.00130 & 0.360 &           0.130 &             0.0356 &                      0.0022 &         2.47 &                0.18 & \tablenotemark{a} &           \cite{Rodriguez2018c}  \\
374180079 &    * K2-266 d &         14.6970000 &             0.0003450 &                2457944.83930 &                     0.00120 & 0.290 &           0.145 &             0.0382 &                      0.0022 &         2.65 &                0.19 & \tablenotemark{a} &           \cite{Rodriguez2018c}  \\
173103335 &        K2-3 c &         24.6467290 &             0.0000430 &                2457329.85688 &                     0.00055 & 0.140 &           0.115 &             0.0266 &                      0.0013 &         1.65 &                0.12 & \tablenotemark{a} &                            \cite{Diamond-Lowe2022} \\
 18310799 &      K2-136 d &         25.5750000 &             0.0023500 &                2457780.81170 &                     0.00650 & 0.677 &           0.045 &             0.0212 &                      0.0013 &         1.42 &                0.11 & \tablenotemark{a} &                                    \cite{Mayo2023} \\
470381900 &    TOI-1696 b &          2.5003110 &             0.0000040 &                2458834.20115 &                     0.00058 & 0.590 &           0.035 &             0.1021 &                      0.0047 &         3.07 &                0.17 & \tablenotemark{b} &                                    \cite{Mori2022} \\
343628284 &  * TOI-1448 b &          8.1122450 &             0.0000180 &                2458713.33750 &                     0.00150 & 0.210 &           0.150 &             0.0663 &                      0.0020 &         2.74 &                0.12 & \tablenotemark{b} &  \cite{Hori2024}  \\
163539739 &    TOI-1278 b &         14.4756700 &             0.0002100 &                2458711.95950 &                     0.00130 & 1.040 &           0.055 &             0.1955 &                      0.0432 &        11.79 &                2.63 & \tablenotemark{b} &                                 \cite{Artigau2021} \\
118327550 &     TOI-244 b &          7.3972250 &             0.0000245 &                2458357.36270 &                     0.00200 & 0.610 &           0.310 &             0.0327 &                      0.0032 &         1.46 &                0.15 & \tablenotemark{b} &                         \cite{Castro-Gonzalez2023} \\
212957629 &  * TOI-2406 b &          3.0766891 &             0.0000024 &                2459115.97600 &                     0.00030 & 0.090 &           0.075 &             0.1285 &                      0.0038 &         2.85 &                0.12 & \tablenotemark{b} & \cite{Hori2024}  \\
 18310799 &      K2-136 b &          7.9752000 &             0.0007900 &                2457817.75630 &                     0.00470 & 0.220 &           0.145 &             0.0137 &                      0.0009 &         0.92 &                0.07 & \tablenotemark{a} &                                    \cite{Mayo2023} \\
243185500 &    TOI-1468 b &          1.8805201 &             0.0000027 &                2458765.67755 &                     0.00089 & 0.465 &           0.071 &             0.0346 &                      0.0017 &         1.39 &                0.08 & \tablenotemark{b} &      \cite{MeierValdes2025, MeierValdes2025errata} \\
150096001 &      K2-133 c &          4.8678400 &             0.0001200 &                2457823.76860 &                     0.00110 & 0.640 &           0.046 &             0.0323 &                      0.0024 &         1.66 &                0.16 & \tablenotemark{a} &                                \cite{Wells2019} \\
423358488 &    * K2-148 c &          6.9226000 &             0.0007000 &                2457387.72777 &                     0.00458 &   - &             - &             0.0251 &                      0.0043 &         1.68 &                0.30 & \tablenotemark{a} &              \cite{Hirano2018b}  \\
374339566 &    * K2-239 b &          5.2400000 &             0.0010000 &                2457908.19100 &                         - & 0.440 &           0.310 &             0.0280 &                      0.0027 &         0.83 &                0.17 & \tablenotemark{a} &          \cite{DiezAlonso2018b}  \\
301289516 &     GJ 9827 d &          6.2018300 &             0.0000030 &                2460265.10196 &                     0.00006 & 0.891 &           0.009 &             0.0313 &                      0.0024 &         2.23 &                0.26 & \tablenotemark{a} &                        \cite{Piaulet-Ghorayeb2024} \\
452866790 &     GJ 3473 b &          1.1980035 &             0.0000019 &                2458492.20408 &                     0.00043 & 0.336 &           0.070 &             0.0318 &                      0.0016 &         1.24 &                0.07 & \tablenotemark{b} &                                  \cite{Kemmer2020} \\
387690507 &     TOI-530 b &          6.3875970 &             0.0000185 &                2458470.19980 &                     0.00165 & 0.330 &           0.095 &             0.1580 &                      0.0144 &         9.15 &                0.87 & \tablenotemark{b} &                                    \cite{Gan2022a} \\
\enddata
\vspace{0.5em}
\tablenotemark{}{Planet parameters for confirmed and validated (denoted with an asterisk in the Planet Name column) planets. Missing parameters that were not reported in the corresponding paper cited in the Planet Source column are denoted with a ``-''. Planet orbital period (P), time of conjunction ($\rm{t_c}$), impact parameter (b) and planet-to-star radius ratio ($\rm{R_p/R_*}$) were taken from the publications cited in the Planet Reference column. \textbf{The updated planet radii ($\rm{R_p}$) were calculated using the stellar radii reported in Table \ref{fig:stellarcomp} in this work, \cite{dressing+17a}, \cite{dressing+19}, \cite{mann+13d}, and \cite{Gore2024}.}}
\tablenotemark{a}{\cite{dressing+19}}
\tablenotemark{b}{\cite{Gore2024}}
\tablenotemark{c}{\cite{dressing+17a}}
\tablenotemark{d}{\cite{mann+13d}}
\end{longdeluxetable}
\end{longrotatetable}
\clearpage

\clearpage
\begin{rotatetable}
\begin{deluxetable*}{llllllllllll}
\tabletypesize{\scriptsize}
\tablewidth{0pt}
\tablecaption{Updated planet parameters for planet candidates in our sample, using stellar parameters reported in Table \ref{tab:updated_params}}
\tablehead{\colhead{TIC}&\colhead{TOI}&\colhead{$\rm{P_{orb}}$}&\colhead{$\rm{\sigma_{P}}$}&\colhead{$\rm{T_{c}}$}&\colhead{$\rm{\sigma_{T_{c}}}$} &\colhead{b} &\colhead{$\rm{\sigma_b}$}&\colhead{$\rm{{R_p}/{R_*}}$}&\colhead{$\rm{\sigma_{{R_p}/{R_*}}}$}&\colhead{$\rm{{R_p}}$} &\colhead{$\rm{\sigma_{R_p}}$} \\ && (d) & (d) &\colhead{(BJD)}& (BJD) &&&&&($\rm{R_\oplus}$)& ($\rm{R_\oplus}$) \label{tab:pcfits}} 
\startdata
415969908 &  233.01 &         11.6700370 &             0.0000120 &                 2458365.2606 &                      0.0020 & 0.803 &         0.019 &              0.051 &                       0.003 &         2.12 &                0.13 \\
415969908 &  233.02 &          7.2011660 &             0.0000240 &                 2458359.4715 &                      0.0060 & 0.598 &         0.073 &              0.036 &                       0.003 &         1.51 &                0.15 \\
312862941 & 1638.01 &          0.9150623 &             0.0000006 &                 2459908.4409 &                      0.0004 &   1.014 &           0.007 &              0.095 &                       0.005 &         7.21 &                0.33 \\
232635922 & 1745.01 &          5.9848710 &             0.0000133 &                 2459653.4738 &                      0.0013 &   0.829 &           0.037 &              0.051 &                       0.003 &         2.44 &                0.16 \\
332477926 & 1754.01 &         16.2143217 &             0.0000468 &                 2459662.8087 &                      0.0018 &   0.564 &           0.060 &              0.036 &                       0.002 &         2.36 &                0.11 \\
138762614 & 1802.01 &         16.7961107 &             0.0193909 &                 2458904.1494 &                      0.0078 &   0.426 &           0.208 &              0.034 &                       0.004 &         2.18 &                0.24 \\
166648874 & 1806.02 &          8.1960758 &             0.0000542 &                 2458905.2286 &                      0.0038 &   1.019 &           0.095 &              0.157 &                       0.076 &         6.72 &                3.93 \\
166648874 & 1806.03 &          0.5019031 &             0.0000023 &                 2458899.7005 &                      0.0026 &   0.863 &           0.049 &              0.040 &                       0.005 &         1.73 &                0.30 \\
321669174 & 2081.02 &          5.1097197 &             0.0000215 &                 2458686.9747 &                      0.0057 &   0.873 &           0.037 &              0.026 &                       0.002 &         1.45 &                0.12 \\
243313296 & 2278.01 &          1.4663344 &             0.0000012 &                 2459690.5881 &                      0.0005 &   0.925 &           0.007 &              0.034 &                       0.002 &         1.21 &                0.10 \\
405763009\tablenotemark{a} & 2433.01 &          - &            - &                 2458746.3710 &                      0.0243 &   1.089 &           0.099 &              0.172 &                       0.083 &         5.83 &                2.84 \\
 59128183 & 2453.01 &          4.4414275 &             0.0000080 &                 2458441.9926 &                      0.0020 &   0.885 &           0.045 &              0.068 &                       0.005 &         3.68 &                0.29 \\
202185707 & 4325.01 &          1.6500673 &             0.0000009 &                 2458469.3689 &                      0.0007 &   0.906 &           0.048 &              0.074 &                       0.006 &         3.63 &                0.35 \\
337385330 & 5112.01 &         15.5328060 &             0.0000730 &                 2460261.2915 &                      0.0028 &   0.678 &           0.081 &              0.027 &                       0.002 &         1.78 &                0.09 \\
389040826 & 5511.01 &          4.7174703 &             0.0159134 &                 2459575.0454 &                      0.0499 &   0.909 &           0.130 &              0.130 &                       0.096 &         7.24 &                6.75 \\
443582629 & 5519.01 &          5.8512587 &             0.0000531 &                 2459284.3922 &                      0.0042 &   0.542 &           0.130 &              0.035 &                       0.003 &         1.39 &                0.14 \\
387974148 & 5551.01 &          2.4054100 &             0.0000331 &                 2459474.7182 &                      0.0035 &   0.604 &           0.143 &              0.030 &                       0.003 &         1.78 &                0.18 \\
422217860 & 5662.01 &          2.3144207 &             0.0000049 &                 2459689.6707 &                      0.0013 &   0.786 &           0.041 &              0.045 &                       0.002 &         2.98 &                0.16 \\
310380289 & 5736.01 &          0.6489868 &             0.0000019 &                 2458738.8948 &                      0.0029 &   0.764 &           0.112 &              0.025 &                       0.003 &         1.55 &                0.14 \\
434116397 & 5955.01 &          0.5902387 &             0.0000017 &                 2459825.8482 &                      0.0015 &   0.776 &           0.051 &              0.022 &                       0.001 &         1.39 &                0.08 \\
262689575 & 5961.01 &          1.6174870 &             0.0000049 &                 2459447.7808 &                      0.0025 &   0.738 &           0.069 &              0.010 &                       0.001 &         0.64 &                0.06 \\
354173595 & 5974.01 &          1.7266943 &             0.0000026 &                 2459447.8998 &                      0.0014 &   0.874 &           0.026 &              0.036 &                       0.002 &         1.49 &                0.08 \\
120045750 & 5981.01 &          7.0142458 &             0.0000310 &                 2459423.5177 &                      0.0022 &   0.645 &           0.084 &              0.039 &                       0.002 &         2.31 &                0.12 \\
284900292 & 5987.01 &         48.7130843 &             0.0003269 &                 2458769.9812 &                      0.0092 &   0.751 &           0.069 &              0.039 &                       0.003 &         2.63 &                0.19 \\
196066560 & 6077.01 &          2.2111104 &             0.0004362 &                 2458713.4706 &                      0.0181 &   0.269 &           0.198 &              0.023 &                       0.003 &         1.64 &                0.24 \\
\enddata
\vspace{0.5em}
\tablenotemark{a}{TESS has only observed two transits of TOI 2433.01. We therefore do not report an orbital period for this planet candidate, and define the time of conjunction ($\rm{T_c}$) as the midpoint of the first observed transit.}
\end{deluxetable*}
\end{rotatetable}
\clearpage

\section{Theoretical Radius Valley Scalings} \label{sec:appendix_theory}

\begin{figure*}[h!]  
    \includegraphics[width=\textwidth]{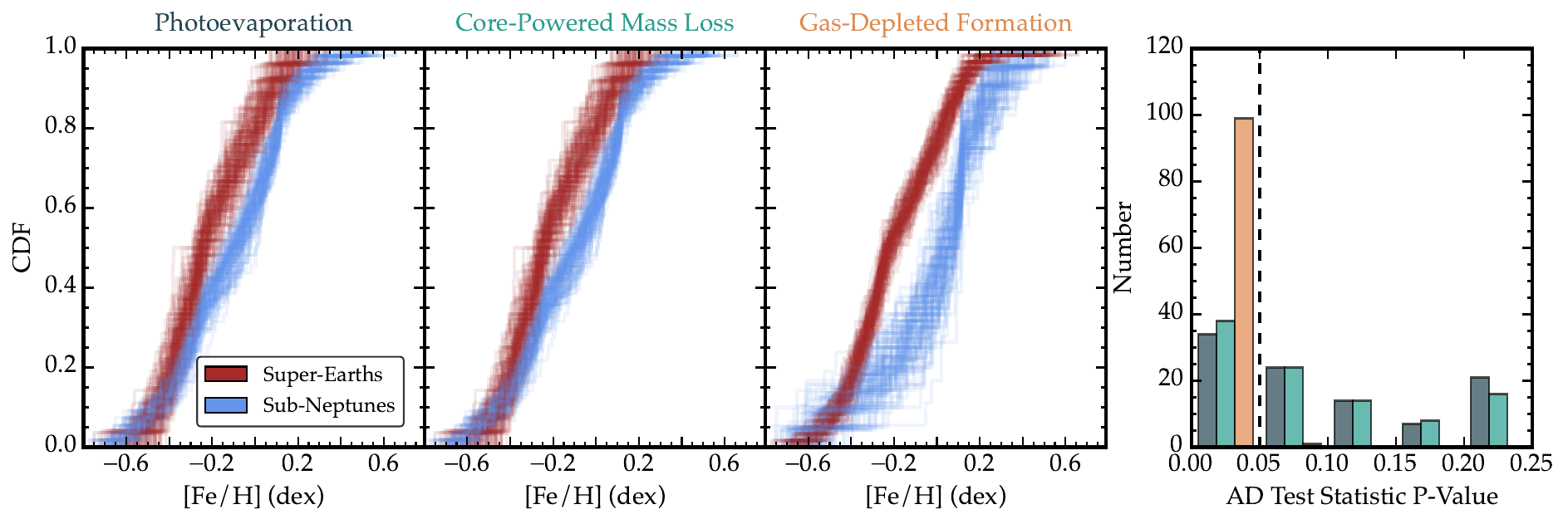}
    \includegraphics[width=\textwidth]{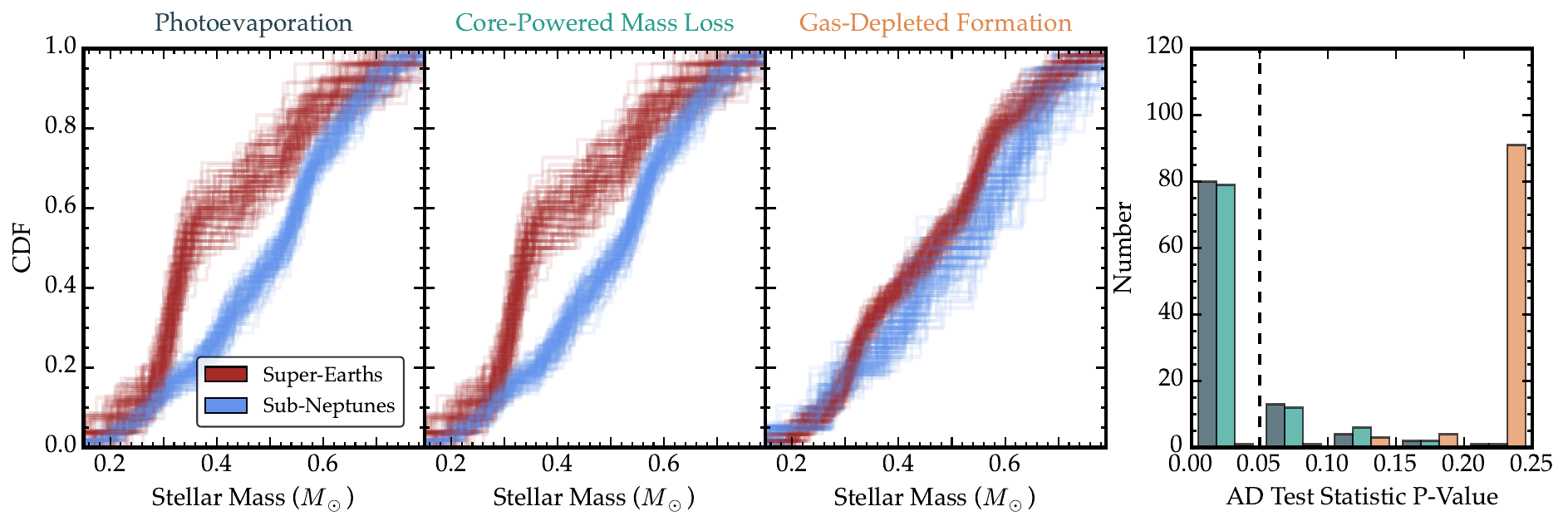}
    \label{fig:theorycdf}
    \caption{(Left 3 plots:) Cumulative distribution function of measured metallicities (top row) and stellar mass (bottom row) for super-Earths (brown) and sub-Neptunes (blue). The distinction between the two populations are set following the radius valley prescriptions for the photoevaporation (navy), core-powered mass loss (teal), and gas-depleted formation (yellow) scenarios. (Right:) Histograms showing the p-values associated with the k-sample Anderson-Darling (AD) \citep{ksamp} tests for the stellar metallicity (top) and mass (bottom) distributions for each of 100 draws from the planet sample plotted in Fig. \ref{fig:pop}. A vertical dashed black line at a p-value of 0.05 indicates a typical upper threshold used to reject the null hypothesis under an AD test. The null hypothesis is that the sub-Neptune and super-Earth host stellar metallicity samples are drawn from the same underlying distribution. This null hypothesis is most inconsistent with the observed stellar metallicity distributions using the gas-depleted formation prescription for the radius valley.}
\end{figure*}

As well as the empirical relations for the radius valley shown in Fig. \ref{fig:pop} and used to generate the super-Earth and sub-Neptune stellar metallicity cumulative distribution functions (CDFs) shown in Fig. \ref{fig:cdf}, we also considered theoretical radius valley scalings. Firstly, we used the scaling presented in \citet{Lopez2018} which was derived assuming photo-evaporative mass-loss due to stellar irradiation. Secondly, we considered a radius valley sculpted by core-powered mass-loss \citep{Gupta2019}. Finally, we considered the results of super-Earth formation in a gas-depleted disk, as presented in \citet{Lopez2018}. Similarly to Fig. \ref{fig:cdf}, we calculated stellar metallicity CDFs for the super-Earth and sub-Neptune planet hosts, as shown in Fig. \ref{fig:theorycdf}. 

We find that the super-Earth and sub-Neptune metallicity samples are inconsistent with being drawn from the same underlying distribution under the gas-depleted formation scaling. Under the photo-evaporative and core-powered mass-loss scenarios, we cannot reject the null hypothesis that the stellar metallicity distributions for super-Earth and sub-Neptune hosts are drawn from the same distribution. 

\citet{Cloutier2020} found that the radius valley around early to mid M dwarfs is most consistent with the gas-depleted scenario in radius-instellation space. Therefore, we argue that the results for the gas-depleted formation radius valley scaling are most appropriate to interpret for our sample, and that these results are in line with the empirical radius valley scaling results discussed in Section \ref{sec:discussion}.

\bibliography{refs}
\bibliographystyle{aasjournal}

\end{document}